\input epsf
\documentstyle[12pt,aasms4]{article}

\def\bib{\par\noindent\hangindent=3mm\hangafter=1}
\def\MH{\rm [M/H]}
\def\mv{M_V}

\def\msol{M_\odot}
\def\mj{M_J}
\def\lsol{L_\odot}
\def\rsol{R_\odot}
\def\zsol{Z_\odot}
\def\mbol{M_{\rm bol}}
\def\mvol{{\rm M}_\odot{\rm pc}^{-3}}
\def\msurf{{\rm M}_\odot{\rm pc}^{-2}}

\def\te{T_{\rm eff}}
\def\gcc{\rm g cm^{-3}}

\def\kms{\rm km\,s^{-1}}
\def\mh{m_{\rm HBMM}}
\def\mz{\rm [M/H]}
\def\simgr{\,\hbox{\hbox{$ > $}\kern -0.8em \lower 1.0ex\hbox{$\sim$}}\,}
\def\simle{\,\hbox{\hbox{$ < $}\kern -0.8em \lower 1.0ex\hbox{$\sim$}}\,}

\begin{document}

\title{THEORY OF LOW-MASS STARS AND SUBSTELLAR OBJECTS}

\author{ Gilles Chabrier and Isabelle Baraffe}
\affil{Centre de Recherche Astrophysique de Lyon (UMR CNRS 5574),\\
Ecole Normale Sup\'erieure de Lyon,
 69364 Lyon Cedex 07, France\\
(chabrier, ibaraffe @ens-lyon.fr)}

\vskip 1.cm

\indent{\bf Short title:} LOW-MASS STARS AND SUBSTELLAR OBJECTS
\vskip 1.cm

\newpage

\begin{abstract}

Since the discovery of the first bona-fide brown dwarfs and extra-solar
planets in 1995, the field of low mass stars and substellar objects
has considerably progressed, both from theoretical and observational viewpoints.
Recent developments in the physics entering the modeling
of these objects have led to significant improvements in the theory and to a
better understanding of their mechanical and thermal properties.
This theory can now be confronted with observations directly in various observational diagrams (color-color, color-magnitude, mass-magnitude, mass-spectral type), a stringent and unavoidable constraint which became possible only recently, with the
generation of synthetic spectra.
In this paper,
we present the current state-of-the-art general theory of low-mass stars and sub-stellar objects, from one solar mass to one Jupiter mass, regarding primarily
their interior structure and evolution. This review is a natural complement to the previous review on the atmosphere of low-mass stars and brown dwarfs (Allard et al 1997). Special attention is devoted to the comparison of the theory with various available observations.
The contribution of low-mass stellar and sub-stellar objects to
the Galactic mass budget is also
analysed.
\end{abstract}

\newpage


{\bf 1 INTRODUCTION}
\dotfill\quad\ 6
\vskip 2pt
\indent Historical Perspective
\vskip 5pt

{\bf 2 THE PHYSICS OF DENSE OBJECTS}
\vskip 2pt
\indent{2.1 Interior Physics}
\dotfill\quad\ 8

\indent \indent {2.1.1 Equation of State}
\dotfill\quad\ 8
\vskip 2pt

\indent \indent {2.1.2 Nuclear rates. Screening factors}
\dotfill\quad\ 11
\vskip 2pt

\indent \indent {2.1.3 Transport Properties}
\dotfill\quad\ 13
\vskip 2pt

\indent {2.2 Atmosphere}
\dotfill\quad\ 16
\vskip 2pt

\indent \indent {2.2.1 Spectral distribution}
\dotfill\quad\ 17
\vskip 2pt

\indent \indent {2.2.2 Transport Properties}
\dotfill\quad\ 22
\vskip 2pt

\indent {2.3 Activity}
\dotfill\quad\ 23
\vskip 2pt
\vskip 5pt

{\bf 3 MECHANICAL AND THERMAL PROPERTIES}
\vskip 2pt

\indent {3.1  Mechanical properties}
\dotfill\quad\ 27
\vskip 2pt

\indent {3.2  Thermal properties}
\dotfill\quad\ 29
\vskip 5pt

{\bf 4 EVOLUTION}
\vskip 2pt
\indent {4.1 Evolutionary tracks}
\dotfill\quad\ 30
\vskip 2pt

\indent {4.2  Mass-magnitude relationship}
\dotfill\quad\ 32
\vskip 2pt

\indent {4.3  Mass-Spectral type relationship}
\dotfill\quad\ 33
\vskip 2pt

\indent {4.4 Irradiated planets}
\dotfill\quad\ 34
\vskip 2pt

\indent {4.5 Color-magnitude diagrams}
\dotfill\quad\
\vskip 2pt

\indent \indent {4.5.1 Pre-main sequence and young clusters}
\dotfill\quad\ 36
\vskip 2pt

\indent \indent {4.5.2 Disk field stars}
\dotfill\quad\ 37
\vskip 2pt

\indent \indent {4.5.3 Halo stars. Globular clusters}
\dotfill\quad\ 39
\vskip 2pt

\indent {4.6 The lithium and deuterium tests}
\dotfill\quad\ 40
\vskip 2pt

\indent {4.7 Low-mass objects in Cataclysmic Variables}
\dotfill\quad\ 41
\vskip 5pt

{\bf 5 GALACTIC IMPLICATIONS}
\vskip 2pt
\indent {5.1 Stellar Luminosity Function and Mass Function}
\dotfill\quad\ 44

\indent {4.2  Brown dwarf Mass Function}
\dotfill\quad\ 47
\vskip 2pt
\vskip 5pt

{\bf 6 CONCLUSIONS}
\dotfill\quad\ 49

\newpage

\section{INTRODUCTION}

Interest in the physics of objects at the bottom of and below the main sequence
(MS) originated in the early demonstration by Kumar (1963) that hydrogen-burning in a stellar core no longer occurs below a certain mass and
that below this limit hydrostatic equilibrium against gravitational collapse
is provided by electron degeneracy. Simple analytical arguments, based on the
balance between the classical ionic thermal pressure and the quantum electronic
pressure yield for this H-burning minimum mass (HBMM) $\mh \approx 0.1\,\msol$, while the first detailed evolutionary
calculations gave $\mh\approx 0.085\,\msol$ (Grossman et al 1974).
Tarter (1975) proposed the term "brown dwarfs" (BD) for objects below this H-burning limit. D'Antona and Mazzitelli (1985) first pointed out that the luminosity below
$\mh$ would stretch by a few orders of magnitude over a few hundredths of
a solar mass, making the observation of BDs a tremendously difficult task.
Subsequent benchmarks in low-mass star (LMS) and BD theory are due primarily to
VandenBerg et al (1983),
D'Antona and Mazzitelli (1985), Dorman et al (1989),
Nelson et al (1986, 1993a) and the Tucson group (Lunine et al 1986, 1989, Burrows et al 1989, 1993). In spite of this substantial theoretical progress, all of these models failed to reproduce the observations at the bottom
of the MS and thus could not provide a reliable determination of the characteristic properties (mass, age, effective temperature, luminosity) of
low-mass stellar and substellar objects. A noticeable breakthrough came from the first
calculations by a few groups of synthetic spectra and atmosphere models characteristic of cool ($\te \simle 4000$ K) objects (Allard 1990, Brett \& Plez 1993, Saumon et al 1994, Allard \&
Hauschildt 1995, Brett 1995, Tsuji
et al 1996, Hauschildt et al 1999). This
allowed the computation of consistent non-grey evolutionary models (Saumon et al 1994, Baraffe et al 1995) and the direct confrontation
of theory and observation in photometric passbands and color-magnitude diagrams,
thus avoiding dubious transformations of observations into theoretical L-$\te$
Hertzsprung-Russell diagrams.

In the meantime, the search for faint (sub)stellar objects has bloomed over the past few years. Several BDs have now been identified since the first discoveries of bona-fide BDs
(Rebolo et al 1995, Oppenheimer et al 1995) either in young clusters
(see Basri, this volume and Mart\'\i n 1999 for reviews, and references therein) or in the  Galactic field (Ruiz et al 1997, Delfosse et al 1997, Kirkpatrick et al 1999a). The steadily
increasing number of identified extra-solar giant planets (EGPs) since the discovery of 51PegB (Mayor \& Queloz 1995, see Marcy \& Butler 1998 for a review) has opened up a new era in astronomy. Ongoing and future ground-based and space-based optical and infrared surveys of unprecedented faintness and precision are likely to reveal hundreds more of red dwarfs, brown dwarfs and giant planets which will necessitate the best possible theoretical foundation. The correct understanding of the physical properties of these objects bears major consequences for a wide range of domains of physics and astrophysics: dense matter physics, planet and star formation and evolution, galactic evolution, missing mass.

A general outline of the basic physics entering the structure and the evolution
of BDs can be found in the excellent reviews of Stevenson (1991)
and Burrows \& Liebert (1993). It is the aim of this review to summarize the
most recent progress realized in the theory of LMS, BDs and EGPs. We will
focus mainly on the internal structure and the evolution of these objects since a comprehensive review on
the atmosphere of LMS and BDs has appeared recently (Allard et al 1997).
We will also consider the implications of LMS and BDs in a more general galactic context and evaluate their
contribution to the Galactic mass budget.

\section{THE PHYSICS OF DENSE OBJECTS}

\subsection{Interior Physics}

\subsubsection{Equation of State}

Central conditions for LMS, hereafter identified generically as objects below a solar mass,
and substellar objects (SSO) for solar composition range from a maximum density $\rho_c\simeq 10^3\,\gcc$ at the hydrogen-burning limit ($m\approx 0.07\,\msol$, see below) to $\rho_c\simeq 10\,\gcc$ for Saturn ($=5\times 10^{-4}\,\msol$) at 5 Gyr, and from $T_c\simeq 10^7$ K for the Sun to $T_c\simeq 10^4$ K for Saturn at the same
age, spanning several orders of magnitudes in mass, density and temperature.
Effective temperatures range from $\sim 6000$ K to $\sim 2000$ K in the stellar domain and extend down to $\sim 100$ K for Saturn.
Molecular hydrogen and other molecules become stable for $kT\simle 3\times 10^{-2} $ Ryd ($T\approx 5\times 10^3$ K),
a condition encountered
in the atmosphere, or even in deeper layers, of most of these objects.
Under these conditions, the interior of LMS and SSO is essentially a
fully ionized H$^+$/He$^{++}$ plasma
characterized by coupling parameters $\Gamma_i=\langle Z^{5/3} \rangle e^2/a_ekT \approx 0.1$-$50$
for classical ions and $r_S=\langle Z \rangle^{-1/3}a/a_0 \approx 0.1-1$ for degenerate electrons.
With  $\Gamma_i=\langle Z^{5/3} \rangle(2.693\times 10^5\,{\rm K}/T)n_{24}^{1/3}$,
$r_S=1.39/(\rho/\mu_e)^{1/3}=1.172\,n_{24}^{-1/3}$,
$n_{24}\equiv n_e/10^{24}$ cm$^{-3}$$\approx (\rho/1.6605\,\gcc)\mu_e^{-1}$
the electron number-density, $a=({3\over 4\pi}{V\over N_i})^{1/3}=a_e\langle Z \rangle^{1/3}$
the mean interionic distance, $\mu_e^{-1}=\langle Z \rangle/\langle A \rangle$ the electron
mean molecular weight, $\rho$ the mass-density, $A$ the atomic mass, $a_0$ the electronic Bohr radius and $<>$ denotes the average number-fraction $\langle X \rangle=\sum_i x_iX_i$ ($x_i=N_i/\sum_i N_i$).
 The temperature is of the order of the electron Fermi temperature $T_F$ so that the degeneracy parameter
$\psi=kT/kT_F\approx 3.314\times 10^{-6}\, T\,  (\mu_e/\rho)^{2/3}$ is of the order
of unity. The
classical (Maxwell-Boltzman) limit corresponds to $\psi \rightarrow + \infty$, whereas $\psi \rightarrow 0$ corresponds
to complete degeneracy. The afore-mentioned thermodynamic conditions yield $\psi \approx 2-0.05$ in the interior of LMS and BDs
implying that finite-temperature effects for the electrons must be included to describe accurately the thermodynamic properties
of the correlated, partially degenerate electron gas. Moreover, the Thomas-Fermi wavelength $\lambda_{TF}= \bigl( kT_F/(6\pi n_e e^2)\bigr)^{1/2}$ is of the order of $a$, so that the electron gas is polarized by the external ionic field and
electron-ion coupling must be taken into account in
the plasma hamiltonian.
Last but not least, the electron average binding energy can be of the order of
the Fermi energy $Ze^2/a_0 \sim \epsilon_F$ so that {\it pressure}-ionization
takes place along the internal profile.
Figure \ref{chabF1.eps} illustrates central characteristic quantities for
LMS and SSOs from the Sun to Jupiter.

Above $\sim 0.4\,\msol$, the structure evolves slowly with increasing mass from a $n$=$3/2$ towards a $n= 3$ polytrope,
which yields the correct $P_c$ for the Sun, due to the growing
central radiative core. This leads to increasing central pressures and densities
for increasing mass (increasing polytropic index) in this mass range.
Below $\sim 0.3$-$0.4\,\msol$, the core becomes entirely convective and follows the behaviour of a $n=3/2$ polytrope. Since the gas is still in the classical regime ($\psi \simgr 1$), $m\propto R$ and the central density increases with decreasing mass, $\rho_c\propto m^{-2}$. Below about the H-burning limit,
electron degeneracy becomes dominant ($\psi \simle 0.1$), so that one approaches the relation $m\propto R^{-3}$ (for $\psi$=0) and density decreases again with decreasing mass, $\rho_c\propto m^2$. These various effects yield a non-monotomic behaviour of the central density and pressure with mass with a minimum in the stellar regime around 0.4 $\msol$ and a maximum near the H-burning limit.

The equation of state (EOS) of low-mass objects thus requires a detailed description of strongly
correlated, polarisable, partially degenerate classical and quantum plasmas, plus an accurate treatment of pressure partial ionization, a severe challenge for dense matter physicists. Several steps towards the derivation of such an accurate EOS for dense
astrophysical objects
have been realized since the pioneering work of Salpeter (1961), with major
contributions by Fontaine et al (1977), Magni and Mazzitelli (1979), Marley and Hubbard (for objects with $m<0.2\,\msol$), H\"ummer \& Mihalas (1988) (primarily devoted to conditions characteristic
of solar-type stellar envelopes) and Saumon and Chabrier (Chabrier 1990, Saumon \& Chabrier
1991, 1992, Saumon et al 1995, hereafter SCVH).
We refer
the reader to Saumon (1994) and SCVH for
a detailed review and an extensive comparison of these different EOS in the domain of interest.
The SCVH EOS presents a consistent treatment of pressure ionization and includes
significant improvements with respect to previous calculations in the treatment of
the correlations in dense plasmas. It compares well with available high-pressure shock-wave experiments in the molecular domain, and with Monte-Carlo simulations
in the fully-ionized, metallic domain (see references above for details).
Recently laser-driven shock-wave experiments on D$_2$ have been conducted at Livermore
(Da Silva et al 1997, Collins et al 1998) which reach for the first time the pressure-dissociation and ionization domain and thus probe directly the thermodynamic properties
of dense hydrogen under conditions
characteristic of BDs and giant planets (GP). The relevance of these experiments for
the interior of SSOs can be grasped from Figure 1 of Collins et al (1998). These experiments have shown the good agreement between the predictions
of the Saumon-Chabrier model and the data (although there is certainly room for improvement). In particular the strong compression
factor arising from hydrogen pressure-dissociation and ionization
observed in the experiment ($\rho/\rho_i \sim 5.8$) agrees well with the predicted theoretical value (see Figure 3 of Collins et al 1998). Any EOS devoted to the description of the interior of dense astrophysical objects must now be confronted with these available data.

The SCVH EOS is a pure hydrogen and helium EOS, based on the so-called additive-volume-law between the pure components (H and He). The accuracy of the additive-volume-law has been examined in detail by Fontaine et al (1977).
The invalidity of this approximation to describe accurately the thermodynamic
properties of the mixture is significant only in the {\it partial ionization} region (see SCVH), which concerns only a few percent of the stellar mass
under LMS and BD conditions.
The
effect of metals on the structure and the evolution of these objects has been examined in detail by Chabrier \& Baraffe (1997, \S 2.1). As shown by these authors, because of their negligible
number-abundance ($\sim 0.2 \%$), metals
do not contribute significantly
to the EOS and barely modify the structure and
evolution ($\sim 1\%$ in $\te$ and $\sim 4\%$ in $L$) of these objects. In the low-density limit characteristic of the
atmosphere, the SCVH EOS recovers the perfect gas limit and thermal contributions from various atomic or molecular species can be added within
the afore-mentioned additive-volume law formalism, which is exact in this regime.
Only in the (denser) metal-depleted atmospheres ($Z \simle 10^{-2} \, Z_\odot$) of the densest objects ($\simle 0.1\,\msol$), one finds
a slight departure from ideality, with a 1\% to 4\%
effect on the adiabatic gradient $\nabla_{ad}$ (Chabrier \& Baraffe 1997).

\subsubsection{Nuclear rates. Screening factors}

Although the complete PP chain is important for nucleosynthesis,
the thermonuclear processes relevant from the energetic viewpoint
under LMS and BD conditions
are given by the PPI chain (see e.g. Burrows \& Liebert 1993, Chabrier \& Baraffe 1997) :

\begin{eqnarray}
p+p\rightarrow  d+e^++\nu_e ;\,
p+d\rightarrow  ^3He\,+\,\gamma ;\,
^3{\rm He}+^3{\rm He}\rightarrow  ^4{\rm He}\,+\, 2p
\end{eqnarray}

Below $\sim 0.7 \msol$, the PPI chain contributes to more than 99\%
of the energy generation on the zero-age Main Sequence (ZAMS), and the PPII
chain to less than 1\%.
The destruction of $^3{\rm He}$ by the above reaction is important only for
T $> 6\times 10^6$ K, i.e masses $m\simgr 0.15\, \msol$ for ages $t<$ 10 Gyr,
for which the lifetime
of this isotope against destruction
becomes eventually smaller than a few Gyr.

As examined in \S4.6, the abundance of light elements ($D,Li,Be,B$) provides a powerful
diagnostic to identify the age and/or the mass of SSOs.
The rates for these reactions (see e.g. Nelson et al 1993b) in the
vacuum, or in an almost perfect gas where kinetic energy largely dominates the
interaction energy, are given by Caughlan and Fowler (1988)
and Ushomirsky et al (1998) for updated values. The reaction rate $R_0$ (in cm$^{-3}$$s^{-1}$) in the  vacuum
is given by $R_0\propto e^{-3 \epsilon_0/kT}$
where $\epsilon_0$ corresponds to the Gamow-peak energy for non-resonant reactions (Clayton 1968).
However, as mentioned in the previous section, non-ideal effects dominate in the
interior of LMS and BDs and lead to polarization of the ionic and electronic
fluids.
These polarization effects due to the surrounding particles
yield an enhancement of the reaction rate, as first recognized by
Schatzman (1948) and Salpeter (1954).
The distribution of particles in the plasma reads :

\begin{eqnarray}
n(r) = \bar n e^{-Ze\phi(r)/kT}\,\,\,\,,\,\,{\rm with} \,\,\,\,\,\,\phi(r) = {Ze\over r} \,+\, \psi(r)
\end{eqnarray}

\noindent where $\psi(r)$ is the induced mean field potential due
to the polarization of the surrounding particles.
This induced potential lowers the Coulomb barrier between the
fusing particles and thus yields an enhanced rate in the
plasma $R=E\times R_0$ where

\begin{eqnarray}
E=lim_{r\rightarrow 0}\, \bigl\{ g_{12}(r) exp({Z_1Z_2e^2\over rkT})\bigr\}
\end{eqnarray}

\noindent is the enhancement (screening) factor and $g_{12}(r)$ the pair-distribution function of particles in the plasma.
Under BD conditions, not only ion screening must be considered
but also electron screening, i.e. $E=E_i\times E_e$.
Both effects
are of the same order ($E_i\sim E_e \sim$ a few) and
must be included in the calculations for
a correct estimate of the light element-depletion factor (see \S 4.6).
Figure \ref{chabF2.eps} portrays the evolution of the central temperature for objects
respectively above, at the limit of and below the hydrogen-burning
minimum mass, with lines indicating the hydrogen, lithium and deuterium
burning temperatures in the plasma.

\subsubsection{Transport Properties}

Energy in the interior of LMS below $\sim 0.4\, \msol$,
 BDs and GPs is transported essentially by convection
(see e.g. Stevenson 1991). According to the mixing length theory (MLT), the convective flux reads (Cox and Giuli 1968) :

\begin{equation}
F_{\rm conv} \propto \rho \, v_{\rm conv}\, C_{\rm P} \, \delta T
\propto
(Q^{1/2}/\Gamma_1^{1/2}) ({l\over H_{\rm P}})^2 \rho C_{\rm P} c_{\rm S} T (\nabla-\nabla_{\rm e})^{3/2}
\end{equation}
\label{conv}

\noindent where $Q = -({\partial {\rm ln} \rho \over \partial {\rm ln} T})_P $ is the volume expansion coefficient, $C_{\rm P}$ is the specific heat
at constant pressure,
$c_{\rm S}= (\Gamma_1P/\rho)^{1/2}$ is the adiabatic speed of sound, $\delta T$ is the temperature excess between the
convective eddy and the surrounding ambient medium,  $\nabla$ the temperature gradient, $v_{\rm conv}$ is the
convective velocity, $l \, {\rm and} \, H_{\rm P}$ denote the mixing length and the
pressure scale height, respectively, and $\nabla_{\rm e} \sim \nabla_{\rm ad}$ is the eddy temperature
 gradient. The last term in parenthesis on the r.h.s. of (4)
defines the fluid superadiabaticity, i.e. the fractional amount by which the real temperature gradient exceeds the adiabatic temperature gradient.
Below a certain mass, the inner radiative core
vanishes and the star becomes entirely convective (VandenBerg et al 1983,
D'Antona \& Mazzitelli 1985, Dorman et al 1989). This minimum mass, calculated with consistent non-grey
atmosphere models and with the most recent OPAL radiative opacities (Iglesias and Rogers 1996) for the
interior, is found to be $m_{conv}=0.35\,\msol$ within the metallicity range
$10^{-2}\le Z/\zsol \le 1$ (Chabrier \& Baraffe 1997 \S3.2). The Rayleigh number is defined as $Ra={gQH_P^3\over \xi \nu}(\nabla-\nabla_{\rm ad})$,
where $g\sim 10^3$-$10^5$ cm s$^{-2}$ is the surface gravity, $\xi$ is the thermal diffusivity (conductive or radiative diffusivity) and $\nu$ is the kinematic viscosity. In the interior of LMS, $Ra\sim 10^{25}$ so that convection is almost perfectly adiabatic
and the MLT provides a fairly reasonable description of this transport mechanism.
Variation of the MLT parameter $\alpha=l/H_P$ between 1 and 2 in the interior is found to be inconsequential below $\sim 0.6\,\msol$ (see Baraffe et al 1997).

However convection can be affected or even inhibited by various mechanisms.
Maximum rotation velocities for LMS and BDs are usually in the range $\sim 20-30\,\kms$
(see e.g Delfosse et al 1998a)
with values as large as 50 $\kms$ and
80 $\kms$ in the extreme case of Kelu 1 (Ruiz et al 1997, Basri et al 2000).
This corresponds to an angular velocity $\Omega\sim 5 \times\, 10^{-4}$
rad ${\rm s}^{-1}$ for a characteristic
radius $R \sim 0.1\,R_\odot$ ($\Omega_{Jup}=1.76\times 10^{-4}$ rad ${\rm s}^{-1}$).
Within most ($> 95\%$ in mass) of the interior of LMS and BDs,
$v_{\rm conv}$ is of the order of $10^2$ cm ${\rm s}^{-1}$ ($<<c_S$) and $H_{\rm P} \sim 10^9$ cm. Thus,
the Rossby number $Ro=v_{\rm conv}/\Omega\,l\approx 10^{-3}$, where $l\sim H_P$, and convection can in principle be inhibited by rotation. However, the fact that lithium
 is not observed in objects with $m\simgr 0.06\,\msol$ (see \S 4.6)
 suggests that some macroscopic transport mechanism (e.g. meridional circulation, turbulence or convection) remains efficient throughout the star. The Reynolds number $Re=v_{conv}l/\nu$ remains largely above unity in stellar and SSO interiors so that convection is not inhibited by viscosity.
Magnetic inhibition requires a magnetic
velocity $v_A=(B^2/4\pi \rho)^{1/2}>v_{conv}$ (Stevenson 1979, 1991), i.e. $B\simgr 10^4$ Gauss under the conditions of interest, only slightly above the predicted fields in these objects (see \S 2.3). However, it has been proposed
by Stevenson
(1979) that in a rapidly rotating fluid with a significant magnetic field,
the Proudman-Taylor theorem no longer applies so that
vertical convective motions could be possible.
Lastly, convection efficiency can be diminished by the presence of a
density gradient.
As shown by Guillot (1995), this may
occur in the interior of gaseous planets, due to a gradient of heavy elements, leading
to inefficient convective transport in the envelope of these objects.

The treatment of convection in the outer (molecular) layers, above and near the photosphere, is a more delicate question. The Rayleigh number in this region is $Ra\sim 10^{15}$, a rather modest value by the usual standards in turbulence. Since the MLT, by definition,
applies to the asymptotic regime $Ra\rightarrow \infty$, it is no longer valid near the photosphere.
Much work has been devoted to the improvement of this formalism
and to the derivation of a non-local treatment of convection. One of the most detailed attempts is due to Canuto and Mazitelli (Canuto \& Mazzitelli 1991, CM). The original CM formalism was
based on a linear stability analysis, whereas energy transport by turbulence is a
strongly non-linear process. It has been improved recently by including
to some extent non-linear modes in the energy rate (Canuto et al 1996) but it yields similar results as the initial CM model. The formalism, however, still requires the calibration of a free parameter, which represents a characteristic "mixing" scale and
in this sense resembles the MLT. A more severe shortcoming of the CM formalism
is that the predicted outermost limit of the convection zone for the Sun and the amount of superadiabaticity in this zone are in significant disagreement, both quantitatively and qualitatively, with the results obtained from coupled
hydrodynamic-radiation
3D simulations (Demarque et al 1999, Nordlund \& Stein 1999). The 3D simulations yield excellent agreement with the observational data at several diagnostic levels: the thermal structure (and the inferred depth of the convection zone), the dynamical structure (vertical velocity amplitude and spectral line synthesis) and the p-mode frequencies and amplitudes for the Sun, without the use of free parameters (see e.g. Spruit \& Nordlund 1990 for a review, and Nordlund \& Stein 1999). This provides now a high degree of confidence in 3-D hydrodynamic models of stellar surface layers and the inferred transition from convective to radiative energy transport.

 The thermal structure, for example, is a very robust property of the numerical models, since it depends relatively little on the level of turbulence (and thus on numerical resolution) (Stein \& Nordlund 2000; Nordlund \& Stein 1999).
These results illustrate the fact that any formalism
based on a homogeneous description
of convection (e.g. MLT, CM) can not describe accurately this strongly inhomogeneous process. Interestingly enough, the standard MLT is found to
compare reasonably well with these simulations, at least for the thermal profile
(see afore-mentioned references),
and thus seems to offer a
reasonable (or least worst!) overall description of convective transport,
even in the small-efficiency convective regions. Indeed, in contrast to 1-D models computed with the CM formalism, the 3-D models do not have a steeper and more narrow superadiabatic structure for the Sun than obtained with the classical B\"ohm-Vitense mixing length recipe (Nordlund \& Stein 1999).
It it thus fair to say that, in the absence of a correct non-local treatment of
convection in LMS and BD interiors, the standard MLT is probably the most reasonable choice, at least for the present objects, LMS and SSOs.
Clearly the developpement of 3-D hydrodynamics models of
the atmosphere of these objects, and the calibration of the mixing length from these simulations, as done for solar-type stars (Ludwig et al 1999), represents one of the next major challenges in
the theory.

Another possible mechanism of energy transport in stellar
interiors is conduction.
Its efficiency
can be estimated as follows:
the distance $l$ over which the temperature changes significantly is
$l\sim (\chi t)^{1/2}$, where $\chi$ is the thermal diffusivity and
$t$ is the time during which the temperature change occurs. For $\chi\sim 10^{-1}$ cm$^2$s$^{-1}$, characteristic of metals, and $t\sim 10^9$ yr, $l\sim 10^2$ km. Heat can thus possibly be transported by conduction only
over a limited range of the interior, providing the density is high enough and
the temperature low enough for electron conductivity to become important.
Indeed, below the HBMM $\sim 0.07 \,\msol$,
the interior becomes degenerate enough during cooling that the
conductive flux $F_{\rm cond}=K_{\rm cond}\nabla T$, where
$K_{\rm cond}= {4ac \over 3} \, {T^3 \over  \rho \kappa_{\rm cond}}$ and $\kappa_{\rm cond}$
is the conductive opacity, becomes larger than the convective flux.
Old enough BDs in the mass-range 0.02-0.07 $\msol$ become degenerate enough to develop a conductive core, which slows down the cooling $L(t)$ (Chabrier et al 2000).

\subsection{Atmosphere}

Knowledge of the atmosphere is needed for two reasons : (i) as a boundary condition for the interior profile in the optically-thick region and (ii) as a description of the emergent radiative flux.
A comprehensive review of the physics of the atmosphere of
LMS and BDs can be found in Allard et al (1997). Only a general outline of
the main
characteristics of these atmospheres will be mentioned in the present review.

\subsubsection{Spectral distribution}

Surface gravity $g=Gm/R^2$ for LMS and SSOs range from $\log g \simeq 4.4$ for a main sequence
1.0 $\msol$ star to $\log g \simeq 3.4$ for Jupiter, with a maximum $\log g \simeq 5.5$ at the H-burning limit, for solar
metallicity.
This yields $P_{ph}\sim g/\bar \kappa \sim 0.1 - 10$ bar and $\rho_{ph}\sim 10^{-6}$-$10^{-4}\,\gcc$ at the
photosphere.
Collision effects become significant under these conditions and
induce molecular dipoles on e.g. H$_2$ or He-H$_2$,
yielding so-called collision-induced absorption
(CIA) between roto-vibrational states of molecules which otherwise would have only quadrupolar transitions (Linsky 1969; Borysow et al 1997). At first order this CIA coefficient scales as $\kappa_{CIA}\propto n_{H_2}^2
\times \kappa_{H_2H_2}$ and thus becomes increasingly important as soon as H$_2$ molecules become stable.
The CIA of H$_2$ suppresses the flux longward of 2 $\mu$m
in the atmosphere of LMS, BDs and GPs. This is one of the reasons (the main one for metal-depleted
objects) for the redistribution of the emergent radiative flux toward shorter
wavelengths in these objects (Lenzuni, Chernoff \& Salpeter 1991; Saumon et al 1994,  Allard \& Hauschildt 1995, Baraffe et al 1997).
The CIA of H$_2$ and H$_2$-He, and the bound-free and free-free opacities of H$^-$ and H$_2^-$
provide the main continuum opacity sources below $\sim 5000$ K for objects with metal-depleted abundances.
Below $\te \simle 4000$ K,
most of the hydrogen is locked into H$_2$ and most of the carbon in CO. Excess oxygen is bound in molecules such as TiO, VO and H$_2$O, with some amounts also in OH and O (see e.g. Figure 1 of Fegley \& Lodders 1996). Metal oxides and metal hydrides (FeH, CaH, MgH) are also present.
The energy distribution of solar-abundance M-dwarfs is thus entirely governed by the line absorption of TiO and VO in the optical, and H$_2$O and CO in the infrared, with no space for a true continuum. The VO and TiO band strength index are used to classify M-dwarf spectral types (see e.g. Kirkpatrick et al 1991). The strongly frequency-dependent absorption coefficient due to the line transitions
of these molecules, as well as Rayleigh scattering ($\propto \nu^4$) shortward of $\sim 0.4\,\mu$m,
yield a strong departure from a black-body energy distribution (see e.g. Figure 5 of Allard et al 1997).

At $\te \approx 2000$ K, near the H-burning limit, signatures of metal oxides and hydrides (TiO, VO, FeH, CaH bands) disappear from the spectral distribution (although some amount of TiO remains present in the atmosphere), as observed e.g. in GD 165B (Kirkpatrick et al 1999b). Alkalies are present under their atomic form. The disappearance of TiO-bands prompted astronomers to suggest a new spectral type classification, the "L"-type,
for objects with $\te$ below the afore-mentioned limit (Mart\'\i n et al 1997, 1998, Kirkpatrick et al 1999a).
Below a local temperature $T\simeq 1300-1500$ K for $P\sim 3-10$ bars, for a solar abundance-distribution,
carbon monoxide CO was predicted to dissociate and the dominant equilibrium form
of carbon was predicted to become CH$_4$ (Allard \& Hauschildt 1995, Tsuji et al 1995, Fegley \& Lodders
1996). Note that this transition occurs gradually with some of the two elements being present in the stability field of the other (see e.g. Fegley \& Lodders
1996, Lodders 1999) so that cool stars could contain some limited amount of methane and CO may be visible in objects with $\te< 1800$ K, as indeed detected in Gl229B (Noll et al 1997; Oppenheimer et al 1998).
This prediction has been confirmed
by the spectroscopic observation of Gliese229B  and the identification of methane
absorption features at 1.7, 2.4 and 3.3 $\mu$m (Oppenheimer et al 1995).
The
presence of methane in its spectrum confirmed unambiguously its sub-stellar nature
(Oppenheimer et al 1995, Allard et al 1996, Marley et al 1996).
Objects like Gl229B, characterized by the strong signature of methane absorption in their spectra, define a new spectral class of objects, the "methane" (sometimes called "T") BDs. Methane absorption in the H and K bands yields a steep spectrum at shorter wavelengths and thus blue near-infrared colors, with $J-K\simle 0$ but $I-J\simgr 5$ (Kirkpatrick et al 1999a, Allard 1999). Although the transition temperature between "L" and "methane" dwarfs is not determined precisely yet, it should lie in the range 1000$\simle \te \simle$ 1700 K (see \S 4.5.2).

Below $\te \sim 2800$ K, complex O-rich compounds
condense in the atmosphere, slightly increasing the carbon:oxygen abundance ratio (see e.g. Tsuji et al 1996, Fegley \& Lodders 1996, Allard et al 1997). Different constituents will condense at a certain location in the atmosphere with
an abundance determined by chemical equilibrium conditions (although non-equilibrium material may form, depending on the time scale of the reactions, as mentioned below)
between the gas phase and the condensed species. The formation of condensed species depletes the gas phase of a number of molecular species (e.g. VO, TiO which will be sequestered into perovskite CaTiO$_3$), modifying significantly the emergent spectrum (see e.g. Fegley \& Lodders 1994 for the condensation chemistry of refractory elements in Jupiter and Saturn). The equilibrium abundances can be
determined from the Gibbs energies of formation either by minimization of the
total Gibbs energy
of the system (Sharp \& Huebner 1990, Burrows \& Sharp 1999), or by computing the equilibrium pressures of each grain species (Grossman 1972, Alexander \& Ferguson 1994, Allard et al 1998).
At each temperature, the fictitious pressure $P_C$ of each condensed phase under consideration is calculated from the partial pressures of the species $i$ which form the condensate (e.g. $Al$ and $O$ for corundum $Al_2O_3$) $P_i=N_i\mathcal{R}T$, where $N_i$ is the number of moles of species $i$ and $\mathcal{R}$ is the gas constant, determined by the vapor phase equilibria. This fictitious pressure $P_C$
is compared to the equilibrium pressure $P_{eq}$, calculated from the Gibbs energy of formation of the condensate. The abundance of a condensed species is determined by the condition that this species be in equilibrium with the surrounding gas phase, $P_C\ge P_{eq}$ (Grossman 1972). The opacities of the grains are calculated from Mie theory.
As suggested by Fegley \& Lodders
(1996) and Allard et al (1998), refractory elements, Al, Ca, Ti, Fe and V are removed from the gaseous atmosphere by grain condensation at about the corundum (Al$_2$O$_3$) or perovskite (CaTiO$_3$) condensation temperature $T\simle 1800$ K.
Rock-forming elements (Mg, Si, Fe) condense as iron and forsterite (Mg$_2$SiO$_4$) or enstatite (MgSiO$_3$) within about the same temperature range (depending on P). Therefore the spectral features of all these elements will
disappear gradually for objects with $\te$ below these temperatures (see e.g. Fegley \& Lodders 1994, 1996, Lodders 1999, Burrows \& Sharp 1999 for detailed calculations).
For jovian-type effective temperatures ($\te \sim 125$ K), H$_2$O and NH$_3$ condense near and below the
photosphere, and water and ammonia bands disappear completely for
$\te \simle 150$ K and 80 K, respectively (Guillot 1999).

The gas abundance strongly depends on pressure and temperature so that the abundances of various species varies significantly with the mass (and $\te$) of the astrophysical body. As shown e.g. in figure 2 of Lodders (1999), the condensation point of the dominant clouds lies much closer to the photosphere for M-dwarfs than for Gl229B or - worse - for Jupiter.
In other words the location of a given grain condensation lies deeper in Jupiter than in Gl229B.
For late M-dwarfs and for massive and/or young BDs, the main cloud formation is predicted to occur very near the photosphere. This is consistent with the fact that all the DENIS and 2MASS objects discovered near the bottom of and below the MS exhibit strong thermal heating
and very red colors (Delfosse et al 1998b, Kirkpatrick et al 1999a)(see \S4.5.2). Indeed the atmospheric heating due to the large grain opacity (the so-called greenhouse or backwarming effect), and the
resulting enhanced H$_2$O dissociation, yield a redistribution of the IR flux,
as proposed initially by Tsuji et al (1996).

A key
question for the grain formation process is the size of the grains (see e.g. Alexander \& Fergusson 1994). The suppression of the flux in the optical in Gl229B suggests
grain sizes of $\sim 0.1\,\mu$m as a source of continuum opacity (Jones \& Tsuji 1997, Griffith et al 1998), although the wings of alkali resonance lines (K I, Na I) is also a source of
absorption in this region (Tsuji et al 1999, Burrows et al 2000). Recent
calculations assume a grain-size distribution in the submicron range (Allard et al 1999). Inclusion of the grain absorption with this size distribution in the atmosphere of objects near the limit of the MS successfully reproduces the observed colors of
GD 165B, Kelu-1 and of DENIS objects (Leggett et al 1998, Allard 1999, Goldman et al 1999, Chabrier et al 2000). However, the spectrum of GL229B shows no indication for dust
in its IR spectrum from 1 to 5 $\mu  m$ (Allard et al 1996, Marley et al 1996, Tsuji et al 1996, Oppenheimer et al 1998, Schultz et al 1998). This suggests
ongoing dynamical processes such as grain settling in SSO atmospheres.
Indeed, as noted by Chabrier et al (2000), although convection occurs only in optically-thick regions ($\tau > 1$) for L-dwarfs, the top of the convection zone lies only about one pressure scale height or less from the photosphere ($\tau \sim 1$) near which most grains condense for objects near the bottom of the MS. This can have important consequences on the formation and settling of atmospheric grains. Indeed, although the temperature at the top of the convective zone is found to be
generally above the condensation temperature
of all grains, convection-induced advection or turbulent diffusion 
could efficiently bring material upward to the region of condensation and maintain small-grain layers, which otherwise will settle gravitationally,

In general, the grain formation process involves a balance between various dynamical timescales, such as the condensation, evaporation, coagulation, coalescence and convection timescales (see e.g. Rossow 1978), not mentioning large-scale hydrodynamical instabilities,
typical of weather conditions on Earth. See for example Marley et al (1999) for an early attempt at modelling BD and EGP atmospheres.
Moreover, although to an excellent first approximation the atmosphere of the solar jovian planets are in thermodynamical equilibrium, non-equilibrium species with chemical equilibrium timescales larger than the convection timescale can be dredged up to the photosphere by convection, as e.g. CO, PH$_3$, GeH$_4$ in Jupiter (Fegley \& Lodders 1994). A correct understanding and a reliable description of this complex grain formation process represent a major challenge for theorists and
the observation of spectra of SSOs
over a significant temperature-range is necessary to provide guidance through this maze.

\subsubsection{Transport Properties}

As mentioned previously, below $\sim 5000$ K, i.e. $m\sim 0.6$-$0.8\,\msol$ depending on the metallicity,
hydrogen atoms recombine, $n_{H_2}$ increases, as does
the opacity
through H$_2$ CIA. The radiative opacity $\kappa$ increases by
several orders of magnitude over a factor 2 in temperature, which in turn decreases the
radiative
transport efficiency (${\mathcal{F}}_{rad} \propto {\nabla  \over \kappa}$). On the other
hand, the presence of molecules increases the number of internal degrees
of freedom (vibration, rotation) and thus the molar specific heat $C_p$, which
in turn decreases the adiabatic gradient $(dT/dP)_{ad}={{\mathcal R}\over C_p}{T\over P}$.
Both effects strongly favor the onset of convection in the
optically-thin ($\tau < 1$) atmospheric layers (Copeland et al 1970, Allard 1990,
Saumon et al 1994, Baraffe et al 1995) with a maximum radial extension for the convection zone around $\te\sim 3000$ K.
The combination of relatively large
densities, opacities and specific heat in layers where H$_2$ molecules
are present
contribute to a very efficient convective transport. Convection is found to be
adiabatic almost up to the top of
the convective region and the extension of the superadiabatic layers
is very small compared to solar type stars
(Brett 1995,  Allard et al 1997).
Thus a variation of the mixing length between $H_{\rm P}$ and 2$H_{\rm P}$ barely
affects the thermal atmospheric profile (Baraffe et al 1997 \S3).

The presence of convection in the optically thin layers
precludes radiative equilibrium in the atmosphere (${\mathcal{F}}_{tot}\ne {\mathcal{F}}_{rad}$) and requires
the resolution of the transfer equation
for radiative-convective atmosphere models
for objects below $\sim 0.7\, \msol$.
This,
and the afore-mentioned strong frequency-dependence of the molecular absorption coefficients,
exclude grey model atmosphere and/or the use of radiative
$T(\tau)$ relationships to
determine the outer boundary condition to the interior structure.
As demonstrated by Saumon et al (1994) and Baraffe et al (1995,
1997, 1998), such
outer boundary conditions overestimate the effective temperature
and the HBMM and yield erroneous $m$-$\te$ and $m$-$L$ relationships,
 two key relations for the calibration of the temperature-scale and for the derivation of mass functions (see
Chabrier \& Baraffe 1997, \S 2.5 for a complete discussion).
Correct evolutionary models for low mass objects require (i) the
connection between the non-grey atmospheric ($P$,$T$) profile, characterized by a given ($\log g, \te$) and the interior
($P$,$T$) profile at a given optical depth $\tau$, preferably large enough for the atmospheric profile to be adiabatic
and (ii) consistency between the atmospheric profiles and the synthetic
spectra used to determined magnitudes and colors.
For a fixed mass and composition,
only one atmosphere profile matches the interior profile
for the afore-mentioned boundary condition. This determines the complete stellar model (mass, radius, luminosity, effective temperature, colors) for
this mass and composition (Chabrier \& Baraffe 1997, Burrows et al 1997 for objects below 1300 K (their figure 1)).

\subsection{Activity}

The {\it Einstein} and {\it Rosat} surveys of the solar neighborhood and of several open clusters have allowed the identification of numerous M-dwarfs as faint X-ray sources, with typical luminosities of the order of the average solar luminosity $L_X\sim 10^{27}$ erg s$^{-1}$, reaching up to $\sim 10^{29}$ erg s$^{-1}$ in young clusters (see e.g. Randich, 1999 and references therein for a recent review). This suggests that late M-dwarfs are as efficient coronal emitters as other cool stars in terms of $L_X/L_{bol}$, which expresses the level of X-ray activity. Moreover about 60\% of M-dwarfs with spectral-type $>M5$ ($\mbol\simeq 12$) show significant chromospheric activity with $\log (L_{H\alpha}/L_{bol})\simeq -3.9$ (Hawley et al 1996). These observations provide important empirical relationships between age, rotation and activity.
The $L_X/L_{bol}$ ratio seems to saturate above a rotational velocity threshold at a limit
$\log(L_X/L_{bol})\simeq -3$, implying that the intrinsic coronal emission of LMS and BDs is quite low, decreases with $L_{bol}$ and thus mass and does not increase with increasing rotation above the threshold limit. This suggests a saturation relation in terms of a Rossby number, i.e. the ratio of the rotational period over the convective turnover time $Ro={P\over t_{conv}}$ (see \S2.1.3). Because $t_{conv}$ increases for lower-mass stars, they will saturate at progressively larger periods, i.e. lower rotational velocities. The velocity threshold for a 0.4 $\msol$ star is estimated around 5-6 km s$^{-1}$ (Stauffer et al 1997). Comparison of the Pleiades and the Hyades samples shows a steep decay of the X-ray activity in the Hyades. This result is at odds with a simple, monotonic rotation-activity relationship. The spin-down timescales for M-dwarfs are much longer than for solar-type stars. Hyades M-dwarfs do indeed show moderate or even rapid rotation ($v \sin i \ge 10$ km s${-1}$) and thus should show strong X-ray emission. The fainter $L_X$ in the Hyades than in the Pleiades thus reinforces the saturation scenario.

Delfosse et al (1998a) obtained projected rotational velocities and fluxes in the H$_\alpha$ and H$_\beta$ lines for a volume-limited sample of 118 field M-dwarfs with spectral-type M0-M6. They found a strong correlation between rotation and both spectral type (measured by $R-I$ colors) and kinematic population: all stars with measurable rotation are later than M3.5 and have kinematic properties typical of the young disk population, or intermediate between young disk and old disk. They interpret this correlation as evidence for a spin-down timescale that increases with decreasing mass, with this timescale being a significant fraction of the age of the young disk ($\sim 3$ Gyr) at spectral type M4 ($\sim 0.15\,\msol$). These data confirm the saturation relation inferred previously for younger or more massive stars: $L_X/L_{bol}$ and $L_{H_\alpha}/L_{bol}$ both correlate with $v\,\sin i$ for $v\,\sin i\simle 4$-5 km s$^{-1}$ and then saturate at 10$^{-2.5}$ and 10$^{-3.5}$, respectively.
Recently, Neuh\"auser and Comeron (1998) have detected X-ray activity in young BDs with $m\approx 0.04\,\msol$ in the Chamaleon star-forming region, with $\log L_X=28.41$ and $\log L_X/L_{bol}=-3.44$. Coronal activity in BDs has been confirmed by Neuh\"auser et al (1999), with all the objects belonging to young clusters or star-forming regions.

All these observations show that there is no drop in $L_X/L_{bol}$
in field stars up to a spectral type of $M7$, which is well below the limit $m\sim 0.3\,\msol$ where stellar interiors are predicted to become fully convective. However, recent observations of the 2MASS objects (Kirkpatrick et al 1999, Fig. 15a) seem to show a significant decrease of activity for L-type objects, which confirms the decline of activity suggested previously for very late M-spectral types ($\simgr M8$)(Hawley et al 1996, Tinney \& Reid 1998). In any event, X-ray emission does not disappear in fully convective stars and even at least young SSOs can support magnetic activity. A fossil field can survive only over a timescale $\tau_d\approx R^2/\eta\sim$ a few years for fully convective stars ($\eta \equiv \eta_t$ is the turbulent magnetic diffusivity and $R$ the stellar radius) so that a dynamo process is necessary to generate the magnetic field. The data suggest that (i) the dynamo-generation believed to be at work in the Sun does not apply for very-low mass objects, (ii) dynamo generation in fully convective stars is as efficient as in stars with a radiative core and (iii) whatever the dynamo mechanism is, it is bounded by the saturation condition at least for late-type stars.

It is generally admitted that magnetic activity in the Sun results from the generation of a large-scale toroidal field by the action of differential rotation on a poloidal field at the interface between the convective envelope and the radiative core, where differential rotation is strongest, known as the tachocline (Spiegel \& Weiss 1980, Spiegel \& Zahn 1992). In this region a sufficiently strong magnetic field is stable against buoyancy. In this case the shear dominates the helicity, this is the so-called $\alpha$$\Omega$ dynamo-generation, which predicts a correlation between activity and rotation, as observed in solar-type stars.

As shown initially by Parker (1975), buoyancy prevents the magnetic field generated by the turbulent motions in a convective zone to become global in character. For this reason, the absence of a radiative core, i.e. of a region of weak buoyancy and strong differential rotation, precludes in principle the generation of a large-scale magnetic field. A dynamo generated by a turbulent velocity field that would generate chaotic magnetic fields in the absence of rotation has been proposed as an alternative process for fully convective stars (Durney et al 1993). They found that a certain level of magnetic activity can be maintained without the generation of a large-scale field. The turbulent velocity field can generate a self-maintained {\it small-scale} magnetic field providing the magnetic Reynolds number $Re=(v l/\eta)^2$ is large enough. The scale of this field is comparable to that of turbulence and in rough energy equipartition with the fluid motions
i.e. ${1\over 8\pi}\langle (B_P^2+B_\phi^2) \rangle\approx {1\over 2}\langle \rho v_{conv}^2\rangle$. Rotation is not essential in this case but it increases the generation rate of the field.

Recently K\"uker and Chabrier (in preparation) explored an other possibility, namely the generation of a {\it large-scale} field by a pure $\alpha^2$-effect. In the $\alpha^2$-dynamo, helicity is generated by the action of the Coriolis force on the convective motions in a rotating, stratified fluid. The $\alpha^2$-effect strongly depends on the Rossby number, or the equivalent Coriolis number $\Omega^\star=2t_{conv}\Omega=4\pi/Ro$. In low-mass objects, the convective turnover time is longer than the rotation period (see \S2.1.3), so that $\Omega^\star>>1$. They find that the $\alpha^2$-dynamo is clearly supercritical. It generates a large-scale non-axisymmetric steady (co-rotating) field that is symmetric with respect to the equatorial plane.
Equipartition of energy yields field strengths of several kiloGauss.
The possible decrease of activity in very late-type stars, as observed for example with the small chromospheric activity of the M9.5 BD candidate BRI 0021-0214 in spite of its fast rotation ($v\,\sin i=$40 km s$^{-1}$) (Tinney et al 1998), is a more delicate issue that requires the inclusion of dissipative processes. Indeed, these calculations show that $\alpha^2$-dynamo can efficiently generate a large-scale magnetic field in the interior of fully convective stars. Since conductivity decreases in the outermost layers of the star, however, there no current would be created by the field in these regions, and thus no dissipative process and no activity (see e.g. Meyer \& Meyer-Hofmeister 1999).

The observed continuous transition in rotation and activity at the fully convective boundary suggests in fact that the $\alpha^2$-dynamo is already at work in the convection zones of the more massive stars. An interesting possibility for verifying the present theory would be Doppler imaging of fast-rotating LMS and BDs. Turbulent dynamo is likely to yield a spatially uniform chromospheric activity whereas the large-scale $\alpha^2$ process suggested by K\"uker \& Chabrier generates asymmetry. Moreover, non-axisymmetric fields can propagate in longitudinal directions without any cyclic variation of the total field energy whereas dynamo waves generated by $\alpha$$\Omega$ processes propagate only along the lines of constant rotation rate (Parker 1955). Therefore we do not expect cycles for uniformly rotating (fully convective) stars.

\section{MECHANICAL AND THERMAL PROPERTIES}

\subsection{Mechanical Properties}

Figure \ref{chabF3.eps} portrays the mass-radius behaviour of LMS and
isolated SSOs from the Sun to Jupiter
for $t=6\times 10^7$ (dash-dot line) and 5$\times 10^9$ yr (solid line)
for $Z=Z_\odot$ and $Z=10^{-2}\times Z_\odot$ (dash-line).
The time required
to reach the ZAMS, arbitrarily defined as the time when $L_{\rm nuc}$ = 95\% $L_{\rm tot}$, for solar metallicity LMS is given in Table \ref{table.zams}.

The general $m$-$R$ behaviour reflects the physical properties characteristic of the interior of these objects, as inferred from Figure \ref{chabF1.eps}.
For $m \simgr 0.2 \,\msol$, $\psi>1$ for all ages so that
the internal pressure is dominated by the classical perfect gas ion+electron contribution ($P=\rho kT/\mu m_H$, neglecting the molecular rotational/vibrational excited level contributions) plus the correcting Debye contribution arising from ion and electron interactions ($P_{DH}\propto -\rho^{3/2}/T^{1/2}$) and $R\propto m$ in first order from hydrostatic equilibrium.
When the density becomes large enough during pre-MS contraction
 so that $\psi < 1$, which occurs
 at t $\simgr$ 50 Myr for $m \, = 0.15\,\msol$ and at t $\simgr$ 10 Myr for the
HBMM $m \, = 0.075\,\msol$,
the EOS starts to be dominated by the contribution
 of the degenerate electron gas ($P\propto \rho^{5/3}$).
This yields a minimum in the $m$-$R$ relationship
$R_{\rm min} \approx 0.08 R_\odot$ for $m\sim$ 0.06-0.07 $\msol$.
Full degeneracy ($\psi\simeq 0$)
 would yield the well-known zero-temperature relationship $R\propto m^{-1/3}$,
 as in white dwarf interiors, but partial degeneracy and the
non-negligible contribution arising from the (classical) ionic Coulomb
pressure (which implies $R\propto m^{+1/3}$) combine to yield a smoother
relation $R=R_0 m^{-1/8}$ at t=5 Gyr, where $R_0\simeq 0.06\,\rsol$ for
$0.01\,\msol\simle m \simle 0.07\,\msol$, i.e. an almost constant radius.
For the age and metallicity of the solar system this radius is of the order of the Jupiter radius $R_J\simeq 0.10\times R_\odot$.
At 5 Gyr, the radius reaches a maximum $R \simeq 0.11\, R_\odot$ for
 $m\simeq 4\times M_{\rm J}$ (where $M_{\rm J}=9.5\times 10^{-4}\,\msol$
is the  Jupiter mass) (Zapolsky \& Salpeter 1969, Hubbard 1994, Saumon et al. 1996).
Below this limit, degeneracy saturates (see Figure \ref{chabF1.eps}) and the classical ionic pressure contribution becomes important enough so that we recover a nearly classical behaviour. Figure \ref{chabF3.eps} also
 displays the astrophysically determined radii of the two eclipsing
 binary systems YY-Gem (Leung \& Schneider 1978)
and CM-Dra (Metcalfe et al 1996), and of the white dwarf companion GD448
(Maxted et al 1998).
The bump on the 60 Myr isochrone near $m \sim 0.01 \msol$ results from
initial deuterium burning (see \S 3.2).

The $m$-$R$ relationship is essentially determined by the EOS and  is
weakly sensitive to the outer boundary condition and the structure of the
 atmosphere, since the latter represents at most (for 1 $\msol$) a few percent of the total radius of
LMSs and SSOs (Dorman et al 1989, Chabrier \& Baraffe 1997).
The effect of metallicity $Z$ on the $m-$R relation remains modest. A decrease in metallicity
yields a slight decrease in the radius at a given age or
at the same stage of nuclear burning (i.e the same H content). Indeed lower metallicity yields a larger $\te$ at a given mass (see \S 3.2). This in turns implies an increase in the nuclear energy production to reach
thermal equilibrium, and thus a larger central temperature
$T_{\rm c}$. Since $R\propto \mu m/T$ from hydrostatic
equilibrium, where $\mu$ is
the mean molecular weight, determined essentially by hydrogen and helium,
a lower-metallicity star contracts more to reach thermal equilibrium.
As shown by Beuermann et al (1998), the variation
of $R$ with $Z$ predicted by the models of Baraffe et al (1998) is in agreement with the radii deduced from
observations.

\subsection{Thermal Properties}

Figure \ref{chabF4.eps} exhibits the mass-effective temperature relationships
for representative ages and metallicities. The arrows
indicate the onset of formation of H$_2$ near the photosphere (Auman 1969,
Copeland et al 1970, Kroupa et al 1990). As discussed in \S2.2.2, molecular
recombination favors convective instability in the atmosphere. Convection yields a smaller T-gradient $(\nabla \sim \nabla_{ad})$ and thus a cooler structure in the deep atmosphere (Brett 1995, Allard \& Hauschildt 1995, Chabrier \& Baraffe 1997 Figure 5). Therefore, a model with atmospheric convection corresponds to a larger $\te$ since the ($P,T)$ interior-atmosphere boundary is fixed for a given mass (\S 2.2.2). Furthermore, the adiabatic gradient in the regions of H$_2$ recombination
decreases to a minimum value $\nabla_{ad}\sim 0.1$,
compared to $\nabla_{\rm ad} \sim 0.4$ for an ideal monoatomic gas (Copeland et al 1970, Saumon et al 1995 Figure 17). Therefore, even if the atmosphere were already convective without H$_2$, molecular recombination yields a flatter inner
temperature gradient in the atmosphere, and thus enhances the former effect, i.e. a larger $\te$ for a given mass.
Formation of H$_2$ in the atmosphere occurs at higher $\te$ for decreasing
metallicity, because of the denser (more transparent) atmosphere
(see Figure \ref{chabF4.eps}).
The relation between the central and effective temperatures for LMS and SSOs can be inferred from Figures \ref{chabF1.eps} and \ref{chabF4.eps} and is commented on by Chabrier and Baraffe (1997 \S 4.2) and Baraffe et al. (1998 \S 2).
As these authors demonstrate, a grey approximation or the use of a radiative $T(\tau)$ relationship
significantly overestimates the effective temperature (from $\sim 50$ to 300 K depending on the atmospheric treatment).

A representative $T_{\rm c}$-$\rho_{\rm c}$ diagram is shown in Figure \ref{chabF5.eps}
(see also Figure 8 of Burrows et al 1997) and
illustrates the different evolutionary paths for LMS and SSOs in this diagram.
The bumps appearing on the isochrones between $10^6$ and
10$^8$ yr and $\log \, T_{\rm c} \sim 5.4$-5.8 result from
initial deuterium burning.
For stars, $T_{\rm c}$ and $\rho_{\rm c}$ always increase with time until
they reach the ZAMS. For BDs,
$T_{\rm c}$ first increases for $\sim 10^7$-$10^9$ yr,
for masses between $\sim 0.01$
and $\sim 0.07 \, \msol$, respectively.
Then, when degeneracy becomes dominant, $T_{\rm c}$ reaches
a maximum and decreases.
For objects with $m\simle 5\times 10^{-3}\,\msol$, T$_c$
always decreases for $t \simgr 1$ Myr.
In the substellar regime there is no steady nuclear energy generation, by definition, and
the evolution is governed by the change of internal energy
$\int_M {d  E \over dt} dm$
and the release of contraction work $\int_M {P\over \rho^2}{d\rho \over dt}dm $.
Even when degeneracy becomes important, SSOs keep contracting, though very slowly,
within a Hubble time.

\section{EVOLUTION}

\subsection{Evolutionary tracks}

Figure \ref{chabF6.eps}
exhibits $\te(t)$ obtained from consistent non-grey calculations for several
masses, for $\mz$\footnote{$\mz=\log (Z/ Z_\odot)$}=0 (helium mass fraction $Y=0.275$) and $\mz=-2.0$ ($Y=0.25$),
respectively. Initial deuterium burning (with an initial mass fraction
$[D]_0=2 \times 10^{-5}$) for masses $m\simgr 0.013\,\msol$ proceeds quickly,
at the early stages of evolution, and lasts about $\sim 10^6$-$10^8$ years.
Objects below this limit are not hot enough to fuse deuterium in their core
(see Figure \ref{chabF2.eps}).
For masses above $0.07\,\msol$ for $[M/H]$=0 and
$0.08\,\msol$ for $[M/H]$=-2.0, the internal energy
provided by nuclear burning quickly balances the gravitational contraction energy,
and after a few Gyr the lowest-mass star reaches
complete thermal equilibrium ($L = \int \epsilon_{\rm nuc} dm$,
 where $\epsilon_{\rm nuc}$ is the nuclear energy rate), for both metallicities.
The lowest mass
for which thermal equilibrium is reached defines the HBMM and the related
hydrogen-burning minimum luminosity HBML. Stars with $m \ge 0.4 \msol$
develop a convective core  near the ZAMS for a relatively short time, depending on the mass
and metallicity, which results in the bumps for 0.6 and 1 $\msol$
at respectively $\sim 10^8$ yr and $\sim 3 \times 10^7$
yr (Chabrier and Baraffe 1997, \S3.2).
The dotted lines portray $\te(t)$
for objects with solar abundance when grain opacity is
taken into account in the atmosphere (see \S4.5.2).
The Mie opacity due to the formation of refractory silicate grains produces
a blanketing effect that lowers the effective temperature and luminosity at
 the edge of the main sequence, an effect first noticed by Lunine et al
 (1989). However as a whole,
grain formation only moderately affects the evolution near and below
the bottom of the main sequence and thus the HBMM and HBML.
Models with grainless atmospheres yield
$m=0.072\,\msol$, $L=5\times 10^{-5}\,\lsol$ and $\te=1700$ K at the H-burning limit,
whereas models with grain
opacity give $m=0.07\msol$, $L= 4\times 10^{-5}\,\lsol$
and $\te = 1600$ K, for solar composition (Chabrier et al 2000).
As shown in Figure \ref{chabF6.eps}, young and massive BDs can have the same  effective temperature (or luminosity) as older very-low-mass stars, a possible source of contamination for the determination of the local stellar luminosity function at the bottom of the MS (see \S 5).

Note the quick decrease of $\te$ (and $L$) with time for objects below the HBMM, $L\propto t^\alpha$ with $\alpha \sim -5/4$
(Stevenson 1991, Burrows et al 1994, 1997), with a small dependence of $\alpha$ on
the presence of grains.
Slightly below $\sim 0.072\,\msol$ (resp. 0.083 $\msol$) for $[M/H]$=0
(resp. $[M/H]$ $\le$ -1),
nuclear ignition still takes place in the central part of the star,
but cannot balance the ongoing gravitational contraction (see Figure \ref{chabF2.eps}). Although these objects are sometimes called "transition objects" we prefer to consider them as
{massive BDs, because strictly speaking they will never reach thermal
equilibrium. Indeed, just below the HBMM, the contributions from the nuclear energy source $\int \epsilon_{\rm nuc} dm$ and the entropy source $\int T{dS\over dt} dm$  are comparable and cooling proceeds at a much slower rate than
mentioned above.
Below about $0.07\,\msol$ (resp. $0.08\,\msol$) for [M/H]=0
(resp. [M/H] $\le$ -0.5), the energetic contribution
arising from hydrogen-burning, though
still present for the most massive objects, is orders of magnitude smaller than the internal heat content, which provides essentially
all the energy of the star ($\epsilon_{\rm nuc} << T|{dS\over dt}|$).
Once on the ZAMS the radius for stars is essentially constant, whereas for BDs
the contraction slows down when $\psi \simle 0.1$ (Figures \ref{chabF1.eps} and \ref{chabF3.eps}).

The effects of metallicity on the atmosphere structure (Brett 1995, Allard and Hauschildt 1995) and on the evolution (Saumon et al 1994, Chabrier and Baraffe 1997) have been discussed extensively in the previously-cited references and
can be apprehended with simple arguments:
the lower the metallicity $Z$, the lower the mean opacity $\bar{\kappa}$ and the more transparent
the atmosphere so that the same
optical depth lies at deeper levels
and thus higher pressure (${dP\over d\tau}={g\over \bar{\kappa}}$).
Therefore for a given mass ($\log g$) the $(T,P)$ interior profile matches, for a given optical depth $\tau$, an atmosphere profile with larger $\te$ (dash-line on Figure \ref{chabF4.eps}),
and thus higher luminosity $L$ since the radius depends only weakly on the metallicity. The consequence is a larger HBMM
for lower $Z$ since a larger $L$
requires more efficient nuclear burning to reach thermal equilibrium, and thus a larger mass.

A standard way to display evolutionary properties is a theoretical
Hertszprung-Russell diagram (HRD). Such a HRD for LMS and SSOs is
shown in Figure \ref{chabF7.eps} for several masses and isochrones from
1 Myr to 5 Gyr, whereas Figure \ref{chabF8.eps} shows evolutionary tracks in a $\log \,g$-$\te$ diagram (see also Burrows et al 1997). These figures allow the determination of the mass and age of an object from
the gravity and effective temperature inferred from its spectrum.

\subsection{Mass-magnitude}

One of the ultimate goals of stellar theory is an accurate determination of the
mass of an object for a given magnitude and/or color.
 Figure \ref{chabF9.eps} (see also Baraffe et al 1998) shows the comparison between theory
and observationally-determined masses in
the $K$-band. The solid line corresponds to a 5$\times 10^9$ yr isochrone, for which the lowest-mass stars have settled on the MS, for a solar metal-abundance ($\MH=0$), whereas the dotted line corresponds to a 10$^8$ yr isochrone for the same
metallicity and the dashed line corresponds to a 10$^{10}$ yr
isochrone for $\MH=-0.5$, representative of the thick-disk population.
 A striking feature is the weak
metallicity-dependence in the K-band,
compared to the strong dependence in the V-band (see Figure 3 of Baraffe et al
1998).
As discussed in this paper, this stems from two different effects.
 On one hand, the increasing opacity in the optical, dominated by TiO and VO lines, and the decreasing H$_2$ opacity in the K-band
with increasing metallicity shift the peak of the flux
toward larger wavelengths. Thus, for fixed $\te$ the V-flux decreases and the K-flux
increases with increasing $\MH$.
On the other hand, for a given mass, the total flux (and $\te$) decreases
 with increasing metallicity, as mentioned in \S4.1. These two effects add up in the V-band and yield an important variation of the V-flux with metallicity.
In the K-band, they cancel and yield similar fluxes for a given mass below $\sim 0.4\,\msol$ ($\te\sim 3500$-$4300$ K, depending on $\mz$) at
 different metallicities.
These arguments remain valid as long as H$_2$ CIA does not
significantly depress the K-band, as it does for very metal-depleted objects.

\subsection{Mass-Spectral type}

The knowledge of spectral type ($Sp$) is extremely useful for analyzing objects
with unknown distance or with colors altered by reddening, as in young
clusters, for example. The determination of a mass-$Sp$ relationship
provides a powerful complement to the mass-magnitude relationship
for assigning a mass to an object.
Based on spectroscopic observations of low-mass nearby composite
systems, Kirkpatrick \& McCarthy (1994) have determined an empirical $m$-$Sp$  relationship for
M-dwarfs, restricted to M2-M6 spectral types.
A theoretical relation has been
derived by Baraffe and Chabrier (1996) from M0 to M10, emphasizing the non-linear behavior of this relationship near the bottom
of the MS, for $>$M6. Figure \ref{chabF10.eps} shows the $m$-$Sp$
relationship for M-, K- and G-dwarfs based on the Baraffe et al (1998) models. The spectral type for K- and G-dwarfs is derived from the model synthetic color ($I-K$) through  the empirical $Sp(I-K)$ relation recently derived
by Beuermann et al (1998).
As stressed by Baraffe and Chabrier (1996), the $m$-$Sp$ relationship depends crucially
on age and metallicity.
At $\sim$ 1 Gyr, objects with $Sp$ later than M10 are
below the HBMM and can be considered as bona-fide BDs. This limit decreases to
$\sim$ M7 at 100 Myr and $\sim$ M6 at 10 Myr. A difficulty arises when dealing with very young objects ($t \simle$ 10 Myr) because of gravity effects.
No reliable $Sp$-color relationship is presently available for these objects
for which spectroscopic and photometric properties are found to be
intermediate between giants and dwarfs (Luhman 1999). The present
$m$-$Sp$ relationships should be extended in the future to dwarfs cooler
than M9.5-M10, the so-called L-dwarfs and methane-dwarfs. Note however that SSOs evolve at different rates through a series of spectral types as they cool, so we cannot associate a given spectral type with a specific mass for these objects.

\subsection{Irradiated planets}

The mass-radius relation and evolutionary sequences described above
reflect the relations for isolated objects. After a rapid initial accretion phase and subsequent hydrodynamical collapse, planets orbiting stars will evolve differently. As shown by Hubbard (1977), illumination from a parent
star will yield thermal expansion of the less massive (low gravity) objects, toward an asymptotic temperature $T_{eq}$ set only by the thermalized photons from the parent star and toward a (larger)
asymptotic equilibrium radius $R_p$, with
$R_p =2a(L_{eq}/L_\star)^{1/2}[1/(1-A)]^{1/2}$ and $T_{eq}=(1-A)^{1/4}(R_\star/2a)^{1/2}T_{eff_\star}$, where $a$ is the orbital distance, $A$ the Bond albedo, $L_\star=4\pi \sigma R_\star^2 T_{eff_\star}$ the parent star luminosity, effective temperature and radius
and $L_{eq}=4\pi \sigma R_p^2 T_{eq}^4$ denotes the equilibrium luminosity (see
Saumon et al 1994). As noted in  Guillot et al (1996) and Guillot (1999),
EGPs in close orbits are heated substantially by their parent star and their atmosphere
cannot cool substantially. They develop an inner radiative region, as a result of this stellar heating and contract at almost
constant $\te$, at a much smaller rate than if they were not heated by the star.
Planets whose luminosity is larger than the absorbed stellar flux of the parent star just evolve along a fully convective Hayashi track for $\sim 10^6$ yr before reaching the afore-mentioned equilibrium state.

The flux from an irradiated planet includes two contributions: the intrinsic thermal emission and the reflected starlight (the Albedo contribution) (see e.g.
Seager \& Sasselov 1998, Marley et al 1999):

\begin{eqnarray}
{\mathcal{F_\nu}} = ({R_p\over d})^2{\mathcal{F}}_\nu^p + ({A\over
4}) P(\phi) ({R_\star \over d})^2 ({R_p \over a})^2 {\mathcal{F}}_\nu^\star
\end{eqnarray}

\noindent where $d$ is the distance of the system to Earth and $P(\phi)$ is the dependence of the reflected light upon the phase angle between the star, the planet and Earth ($P=1$ if the light reflected by the planet is redistributed uniformly over 4$\pi$ steradians). Figure \ref{chabF11.eps} illustrates the flux from a young EGP (assuming the irradiation is isotropic) with $\te=1000$ K orbiting a Sun-like star at 15 pc for various orbital distances.

\subsection{Color-magnitude diagrams}

Thanks to the recent progress both on the observational side, with the development of
 several ground-based and space-based observational surveys of unprecedented sensitivity
 in various optical and infrared passbands, and on the theoretical side, with the
 derivation
of synthetic spectra and consistent evolutionary calculations based on non-grey
atmosphere models, it is now possible to confront theory with observation directly in
the observational planes, i.e. in color-color and color-magnitude diagrams (CMD).
This avoids dubious color-$\te$ or color-$M_{bol}$ transformations and allows more accurate
determinations of the intrinsic properties ($m,L,\te,R$) of an object from its
 observed magnitude and/or color.
In this section we examine the behaviour of LMS and SSOs in various CMDs
 characteristic of different populations in terms of age and metallicity.
These diagrams capture the essence of the observational signatures of the
very mechanical and thermal properties of these objects, described in the
 previous sections.

\subsubsection{Pre-main sequence and young clusters}

Numerous surveys devoted to the detection of SSOs have been conducted
in young clusters with ages spanning from $\sim$ 1-10 Myr to $\simgr 10^8$ yr for the Pleiades or the Hyades (see Mart\'\i n 1999 and Basri this volume for
reviews).
Observations of young clusters present two important advantages,
namely (i) all objects in the cluster are likely to be coeval within a reasonable
range, except possibly in star forming regions, where the spread in ages for
cluster members can be comparable with the age of the cluster, (ii) young objects are brighter for a given mass (see Figure \ref{chabF6.eps}),
which makes the detection of very-low mass objects easier.
 Conversely,
they present four major difficulties, (i) extinction caused by the surrounding dust
modifies both the intrinsic magnitude and the colors of the object,
(ii) accurate proper motion measurements are necessary to assess whether the object belongs
to the cluster, (iii) gravity affects both the spectrum and the evolution, (iv) the evolution and
spectrum of
 very young objects ($t \simle$ 1 Myr) may still be affected by the presence
of an accretion disk or circumstellar material residual from the protostellar stage.
Young multiple systems
remove some of these difficulties
and provide excellent tests for
models at young ages.
Like for example the quadruple system GG TAU (White et al 1999),
with components covering the whole mass-range of LMS and BDs from 1 $\msol$
to $\sim$ 0.02 $\msol$.
Models based on non-grey atmospheres (Baraffe et al 1998) are
 the only ones consistent with these observations
(White et al 1999, Luhman 1999). The comparison of $\sim 10^6$ yr isochrones for SSOs with observations in a near-IR CMD is shown in Zapatero Osorio et al (1999)  for the young cluster $\sigma$-Orionis (their figure 1).

\subsubsection{Disk field stars}

Figures \ref{chabF12.eps} and \ref{chabF13.eps} display the CMDs for LMS and SSOs in the optical and the
infrared for various ages and metallicities (see
also Burrows et al 1997). Various sets of models correspond to calculations where (i) grain formation is included in the atmosphere EOS but not in the transfer equations, mimicking a rapid settling below the photosphere (COND models), (ii) grain formation is included both in the EOS and the opacity (DUSTY models) (see Allard et al 1999), (iii) grainless models. Models (i) and (ii) represent extreme cases that bracket the complex grain processes at play in these atmospheres (see \S2.2.1).
The optical sequence shows a monotonic magnitude-color behaviour with 3 changes of slope around $M_{\rm V} \sim$ 10, 13 and 19 for $Z=Z_\odot$, respectively, which correspond
to 0.5 $\msol$, $\te$=3600 K; 0.2 $\msol$, $\te$=3300 K and 0.08 $\msol$,
$\te$ = 2330 K on the 5 Gyr isochrone. They reflect, respectively, the onset of convection in the atmosphere, degeneracy in the core and grain absorption near the photosphere, as described in the previous sections. The latter feature, which yields a steep increase of $M_{\rm V}$ at $(V-I) \sim 5$ does not appear in
dust-free models (cf. Baraffe et  al 1998), where TiO is not depleted by grain formation in the atmosphere and thus absorbs the flux in the optical. Interestingly enough, this steepening is observed
in the sample collected recently by Dahn et al (1999) which includes some DENIS and 2MASS L-dwarf parallaxes. This effect is also observed in the $(R-I)$ CMD of the Pleiades (Bouvier et al 1998), in agreement with the theoretical predictions (Chabrier et al 2000). This clearly illustrates the effect of grain formation in the atmosphere which affects the spectra of LMS and SSOs below $\te \simle 2400$ K.
In the near-infrared color ($J$-$K$) (Figure \ref{chabF13.eps}), grain condensation yields very red colors for the DUSTY models, whereas
COND and grainless models loop back to short wavelengths (blueward) below $m\sim 0.11\msol$, $M_K\sim 10$ for
$\mz$ = -2, and
$m\sim 0.072\msol$ (resp. $\sim 0.04\msol$) at
5 Gyr (resp. 100 Myr) for solar metallicity, i.e. $M_K\sim 12$.
As described in \S2.2, this stems
from
(i) the onset of H$_2$ CIA for low-metallicity objects and (ii)
the formation
of CH$_4$ at the expense of CO when the peak of
the Planck function
moves in the wavelength range characteristic of the absorption bands of this
species, for solar-like abundances.

In terms of colors there is a competing effect between grain and molecular opacity sources for objects at the bottom and just below the MS.
The backwarming effect resulting from the large grain opacity
destroys molecules such as e.g H$_2$O, one of the main sources of absorption in the near-IR
(cf. \S 2.2.1), and
yields the severe reddening near the
bottom of the MS ($\te \simle 2300-2400$ K), as illustrated by the DENIS
(Tinney et al 1998) and 2MASS (Kirkpatrick et al 1999a) objects and by GD165B (Kirkpatrick et al 1999b).
As the temperature decreases, grains condense and settle at deeper layers below the photosphere while methane absorption in the IR increases (see \S 2.2) so that the peak of the flux will move back to shorter wavelengths
and the color
sequence will go from the DUSTY one to the COND one. The exact temperature at which this occurs is still uncertain because it involves the complex grain thermochemistry and dynamics outlined in \S2.2, but it should lie between $\te\sim 1000$ K, for Gl229B, and
$\sim 1700$ K, corresponding to $J$-$K \sim 2$ for
the presently reddest observed L-dwarf with determined parallax LHS102B (Goldman et al 1999, Basri et al 2000).

In spite of the afore-mentioned strong
absorption in the IR,
BDs around 1500 K radiate nearly 90\% (99\% with dust) of their
energy at wavelengths longward of $1 \, \mu$m  and infrared
colors are still preferred to optical colors (at least for solar metal abundance), with $J,Z,H$ as the favored bands for detection, and M$_M\sim$M$_L^\prime\sim 10-11$, M$_K\sim$M$_J\sim 11-12$ at the H-burning limit, at 1 Gyr (see Chabrier et al 2000).

\subsubsection{Halo stars. Globular clusters}

Observations of LMS belonging to the stellar halo population of our Galaxy (also called "spheroid" to differentiate it from the $\rho(r)\propto 1/r^2$ dark halo) or to globular clusters down to the bottom of the MS have been rendered possible with the HST
optical (WFPC2) and IR (NICMOS) cameras and with parallax surveys at
very faint magnitudes (Leggett 1992,
Monet et al 1992, Dahn et al 1995). This
provides stringent constraints for our understanding of LMS structure and evolution for
 metal-depleted objects. The reliability of the LMS theory outlined
in \S 2 has been assessed by the successful confrontation to various
observed sequences of globular clusters ranging
from $\MH \simeq -2.0$ to -1.0\footnote{Globular clusters with an observed $[Fe/H]$ must
be compared with theoretical models with the corresponding metallicity $[M/H]$ = $[Fe/H]$
+ $[O/Fe]$, in order to take into account the enrichment of $\alpha$-elements (see Baraffe et al 1997)} (Baraffe et al 1997).
Figure 8 of Baraffe et al (1997)
portrays several tracks for different metallicities, corresponding to the MS of
various globular clusters. The tracks are
superposed with the subdwarf sequence of Monet et al (1992), which is identified as stellar halo objects
from their kinematic properties.
As shown in this figure,
the pronounced variations
of the slope in the CMD around $\sim 0.5\,\msol$ and $\sim 0.2\,\msol$, which stem from
the very physical properties of the stars, namely the
onset of molecular absorption in the atmosphere and of degeneracy in the
core (see \S 2), are well reproduced by the theory at the correct magnitudes and
colors.
The predicted blue loop in IR colors caused by the ongoing
CIA absorption of H$_2$ for the lowest-mass (coolest) objects (see \S2.2.1
and Figure \ref{chabF13.eps} for $\mz=-2$), has also been confirmed by observations with
the NICMOS camera (Pulone et al 1998).

\subsection{The lithium and deuterium tests}

One ironclad certification of a BD is a demonstration that
hydrogen fusion has not occurred in its core.
As Figure \ref{chabF2.eps} illustrates, lithium burning through the Li$^7(p,\alpha)$He$^4$
reaction occurs at a lower temperature
than is required for hydrogen fusion. The timescale for the destruction of lithium in the lowest-mass stars is about 100 Myr.
Moreover the evolutionary timescale in LMS/BDs is many orders of magnitude larger than the convective timescale (see e.g. Bildsten et al 1997), so core abundances and atmospheric abundances can be assumed identical.
Therefore, the retention of lithium in a fully mixed object older than 10$^8$ yr
signifies the lack of hydrogen burning. This provides the basis of the so-called
"lithium-test" first proposed by Rebolo et al (1992) to identify bona-fide BDs. Evolutionary models based on non-grey atmospheres and the screening factors described in \S 2.1.2 yield a lithium-burning minimum mass $m_{Li}\simeq 0.06\,\msol$
(Chabrier et al 1996). Note from Figure \ref{chabF2.eps} the strong age dependence of the lithium-test : young stars
at $t\simle 10^8$ yr (depending on the mass) will exhibit lithium
(which in passing precludes the application of the lithium-test for the identification of SSOs in star forming regions),
whereas massive BDs within the mass range [0.06-0.07 $\msol$]
older than $\sim 10^8$ yr will have burned lithium.

By observing a young cluster older than $\sim 10^8$ yr,
one can look for the
boundary-luminosity below which lithium has not yet been depleted.
Objects fainter than this limit will definitely be in the substellar domain.
Conversely, the
determination of the lithium depletion edge
yields the age of the cluster, as was first proposed by Basri et al (1996) (see also e.g. Stauffer et al 1998).
Indeed, the lithium test provides a
``nuclear'' age that may be even more powerful a diagnostic than the conventional
nuclear age from the upper main sequence turnoff, since the evolution of the
present fully convective low-mass objects does not depend on ill-constrained
parameters such as mixing length or overshooting.
As noted by Stauffer et al (1998) the age-scale for open clusters based on the (more reliable) lithium depletion boundary has important implications for stellar evolution: the ages of several clusters (Pleiades, $\alpha$-Per, IC 2391) are all consistent with a small but non-zero amount of overshooting to be included in the evolutionary models at the turn-off mass
in order to yield similar ages. See Basri (this issue) for a more detailed discussion of the lithium-test.

The deuterium-test can be used in a similar manner, extending the lithium test
to smaller masses and younger ages, typically $t \simle 10^7$ yr (B\'ejar et al 1999). The $D$-burning minimum mass is predicted as $m_{D}\simeq 0.013 \msol$ (see Figure \ref{chabF2.eps})
and has
been proposed (Shu et al 1987) as playing a key
role in the formation of isolated star-like objects, in contrast to objects formed
in a protoplanetary disk. Deuterium-depletion is illustrated in Figure \ref{chabF14.eps}.
This figure displays the evolution of SSOs above $m_D$
in the I-band.
The left and right diagonal solid lines correspond to
50\% D-depletion ($[D]/[D]_0=1/2$) and 99\% D-depletion, respectively
($[D]_0=2\times10^{-5}$ is the initial mass fraction). The inset
displays the corresponding curves in $\te$.
 Spectroscopic signatures of deuterium include absorption
lines of deuterated water
 HDO between 1.2 and 2.1 $\mu$m (Toth 1997).
The observation of deuterated methane CH$_3$D can not be used
for the deuterium test as an age-indicator:
at the temperature methane forms ($\simle 1800$ K), SSOs above the D-burning minimum mass are old enough for all
 deuterium to have been burned (see Figure \ref{chabF6.eps}).

\subsection{Low-mass objects in cataclysmic variables}

The study of cataclysmic variables (CVs) is closely related to the
theory of LMS and SSOs regarding their inner structure and the
magnetic field generation.
CVs are semi-detached binaries with a white
dwarf (WD) primary and a low-mass stellar or sub-stellar companion
($m \, \simle \, 1 \, \msol$) that
transfers mass to the WD through Roche-lobe overflow
(see e.g. King 1988 for a review). In the current standard model, mass transfer is driven
by angular momentum loss caused by magnetic wind braking for
predominantly radiative stars and by gravitational wave emission for fully
convective objects.
The Roche lobe filling secondary's mean density
$\bar{\rho}$ determines almost entirely the orbital period $P$, with $P_h=k/\bar{\rho}^{1/2}$ (where $k \simeq 8.85$ is a weak
function of the mass ratio, $P_h$ is the orbital period in hr, and $\bar{\rho}=(m_2/R_2^3)/(\msol/\rsol^3)$
is the mean density of the secondary in solar units) (see King
1988). For objects obeying classical mechanics, i.e. in the stellar regime, $m\propto R$ (see \S3.1 and Figure \ref{chabF3.eps}) so that the mean density increases with decreasing mass ($\bar{\rho}\propto 1/m^2$) and the orbital period decreases along evolution as the secondary loses mass $P_h\propto 1/\bar{\rho}^{1/2}\propto m_2$. The situation reverses near and below the HBMM,
when electron degeneracy dominates. In that case, the radius increases very slightly with decreasing mass (Figure \ref{chabF3.eps}), so that the mean density is essentially proportional to the mass ($\bar{\rho}\propto m$) and the orbital period increases along evolution $P_h\propto \sim m_2^{-1/2}$.
Then the secular evolution of the orbital period reverses when the donor becomes a BD.
The analysis of CV orbital evolution thus provides important constraints on LMS and BD internal structure.

The major puzzle of CVs is the observed distribution of orbital
periods, namely the dearth of systems in the 2-3 h period range
and the minimum period $P_{\rm min}$ at 80 min (see King 1988). The most
popular explanation for the period gap is the disrupted magnetic
braking scenario
(Rappaport et al 1983, Spruit \& Ritter 1983), and most
evolutionary models including this process do reproduce the observed period distribution.
However, the recent progress realized
in the field of LMS and SSOs now allows the confrontation of the theoretical
predictions with the observed atmospheric properties (colors,
spectral types) of the secondaries (Beuermann et al 1998, Clemens et al 1998, Kolb \& Baraffe
1999a).
The afore-mentioned standard period-gap model may be in conflict with the observed spectral
type of some CV secondaries (Beuermann et al 1998, Kolb \& Baraffe
1999b).
Although alternative explanations exist for the period gap, none of them has
been proved as successful as the disrupted magnetic braking scenario.
The most recent alternative suggestion, based on a
characteristic feature of the mass-radius relationship of
LMS (Clemens et al 1998), has been shown to fail to reproduce the observed period distribution around the period gap (Kolb et al 1998).
On the other hand,
as discussed in \S2.3, no
change in the level of activity is observed in isolated M-dwarfs along the
transition between partially and fully convective structures at $m \sim 0.35 \msol$
and $Sp\sim$ M2-M4, which indicates that a magnetic field is still generated in the mass-range corresponding to the CV-gap. In this case,
the abrupt decrease of
angular momentum losses by magnetic braking at the upper edge of the period
gap could result from a
rearrangement of the magnetic field when stars become fully
convective, without necessarily implying
 a sudden decline in the magnetic activity, as suggested by Taam \& Spruit (1989,
see also Spruit 1994).

Finally, a $\sim$ 10\% discrepancy still remains between the theoretical and the observed minimum period $P_{\rm min}=$80 min, even when including the most recent
improvements in stellar physics (see Kolb \& Baraffe 1999a for details).
Residual shortcomings in the theory, either in
the EOS or in the atmosphere, cannot be ruled out as the cause of this
 discrepancy. Alternatively, an additional driving mechanism to gravitational
 radiation in fully convective objects can reconcile
predicted and observed $P_{\rm min}$ (Kolb \& Baraffe 1999a).
Because magnetic
activity is still observed in fully convective late spectral type M-dwarfs, magnetic braking could operate in CV secondaries
even down to $P_{\rm min}$, but with a weaker efficiency than above the
period gap.
These open questions certainly require a better understanding of magnetic field generation
and dissipation in LMS and BDs, a point already stressed in \S2.3.

\section{GALACTIC IMPLICATIONS}

\subsection{Stellar Luminosity Function and Mass Function}

It is now well established that visible stars are not numerous enough to account
for the dynamics of our Galaxy, the so-called galactic
dark matter problem. A precise determination of their density requires the correct knowledge of the luminosity function (LF)
down to the H-burning limit, and a correct transformation into a mass-function (MF). The latter issue has been improved significantly with models that describe more accurately the color-color and color-magnitude diagrams of LMS and BDs for various metallicities, and most importantly that provide mass-magnitude relationships in good agreement with the observations (see \S 4). The former issue, however, is not completely settled at present and significant differences still
exist between various determinations of the nearby LF and between the nearby LF and the photometric LF determined from the ground and with the HST, as shown later in this section.

The main problem with the nearby sample is the limited number of stars at faint magnitudes. Although the sample is complete to 20 pc for stars with $\mv<9.5$, it is severely incomplete beyond 5 pc for $\mv > 12$ (Henry et al 1997).
Kroupa (unpublished) derived a nearby LF $\Phi_{near}$ by combining Hipparcos parallax data, which is essentialy complete for $\mv < 12$ at r=10 pc, and the sample of nearby stars (Dahn et al 1986) with ground-based parallaxes for $V>12$ to a completeness distance r=5.2 pc.
The nearby LF determined by Reid and Gizis (1997) is based on a volume sample within 8 pc. Most of the stars in this survey have parallaxes. For all the late K and M dwarfs, however, trigonometric parallaxes are not available and these authors use a spectroscopic TiO-index vs $\mv^{TiO}$ relation to estimate the distance (Reid et al 1995). This sample was revised recently with Hipparcos measurements and new binary detections in the solar neighborhood (Delfosse et al 1999a) and leads to a revised northern 8-pc catalogue and nearby LF (Reid et al 1999). These authors argue that their sample should be essentially complete for $\mv<14$. However, the analysis of completeness limits by Henry et al (1997, their figure 1) shows that the known stellar census becomes substantially incomplete for distances larger than 5 pc. About 35\% of the systems in the Reid et al sample are multiple and $\sim 45\%$ of all stars have a companion in binary or multiple systems.

The photometric LFs, $\Phi_{phot}$, based on large observed volumes via deep pencil-beam surveys, avoid the limitation due to small statistics but introduce other problems such as Malmquist bias and unresolved binaries (see e.g. Kroupa et al 1993, Kroupa 1995),
yielding only the determination of the stellar {\it system} LF. For the HST LF (Gould et al 1998), with $I\simle 24$, the Malmquist bias is negligible because all stars down to $\sim 0.1\,\msol$ are seen through to the edge of the thick disk. A major caveat of photometric LFs, however, is that the determination of the distance relies on a photometric determination from a color-magnitude diagram. In principle this requires the determination of the metallicity of the stars, since colors depend on metallicity (see Figure \ref{chabF12.eps}).
The illustration of the disagreement between the three afore-mentioned LFs, $\Phi_{5pc}$, $\Phi_{8pc}$ and $\Phi_{HST}$, at faint magnitude is apparent in Figure \ref{chabF15.eps}. The disagreement between the two nearby LFs, in particular, presently has no robust explanation. It might come from incompleteness of the 8 pc sample at faint magnitudes. Note also that in the spectroscopic relation used to estimate the distance, $\mv$ can be uncertain by $\sim 1$ mag (see Figure 3 of Reid et al 1995), certainly a source of Malmquist bias, even though the number of stars without parallax in this sample is small ($\sim 10$\%) and a Malmquist bias on this part should not drastically affect the results. On the other hand, as noted in \S4.1, the end of the 5-pc sample might be contaminated by a statistically significant number of young BDs, or by stars with slightly under-solar abundances.

Figure \ref{chabF16.eps} compares the MFs obtained from the two afore-mentioned nearby LFs, using the mass-magnitude relations described in \S 4.2. In practice, the observed sample constitutes a mixture of ages and metallicities. However the
Hipparcos CMD indicates that $\sim 90\%$ of disk stars have $-0.3<\MH<0.1$ (Reid 1999) so the spread of metallicity is unlikely to significantly affect the derivation of the MF through the mass-magnitude relation. Moreover the choice of $m$-M$_K$ minimizes metallicity effects (see Figure \ref{chabF9.eps} and figure 3 of Baraffe et al 1998). The MS lifetimes for stars with $m\simle 1\,\msol$ are longer than a Hubble time and age variations in the mass-magnitude relation will affect only objects
$\simle 0.15\,\msol$ younger than $\sim 0.5$ Gyr (see Figure \ref{chabF9.eps}).

Superposed is a power-law MF normalized at 0.8 $\msol$ on the Hipparcos sample $dn/dm=0.031(m/0.8)^{-\alpha}=0.024\,m^{-\alpha}\,\msol^{-1}pc^{-3}$ with $\alpha=1.2$. As shown in Figure \ref{chabF16.eps}, this provides a very reasonable description of the low-mass part of the MF, within a variation of $\sim 10\%$ for $\alpha$, thus confirming the previous analysis of Kroupa et al (1993), except for the very last bin obtained from Kroupa's LF, which predits about twice as many stars at 0.1 $\msol$ (out of the figure). Similar results are obtained from the bolometric and $M_K$- LFs for the 8-pc sample. Integration of this MF yields very-low-mass stars ($m\le 0.8\,\msol$) number- and mass-densities $n_{VLMS}\simeq 7.0\times 10^{-2}$ pc$^{-3}$ and $\rho_{VLMS}\approx 2.0\times 10^{-2}$ $\mvol$, respectively.
Adding up the contribution from more massive stars, $\rho_{>0.8}\approx 1.4\times 10^{-2}$ $\mvol$ (Miller \& Scalo 1978) and stellar remnants, $\rho_{WD+NS}\approx 3.0\times 10^{-3}$ $\mvol$ yields a stellar density $\rho_\star \approx 3.4
\times 10^{-2}$ $\mvol$ in the Galactic disk, i.e. a surface density $\Sigma_\star\approx 22\pm 2$ $\msurf$ (assuming a scale height $h= 320$ pc).
A Salpeter MF, $dn/dm\propto m^{-2.35}$, all the way down to the bottom of the MS would overestimate the density at 0.1 $\msol$ by more than a factor of 10.

For the Galactic spheroid the question is more settled. The LFs determined from nearby surveys (Dahn et al 1995) and from the HST (Gould et al 1998) are comparable within about 2 sigmas, and yield a MF with $\alpha \simle 1$ (Graff \& Freese 1996, Chabrier \& M\'era 1997, Gould et al 1998) and a stellar
number- and mass-density $n_\star < 10^{-3}\, pc^{-3}$ and $\rho_\star < 4.0\times 10^{-5}$ $\mvol$,  i.e. an optical depth $\tau\sim 10^{-9}$, about 1\% of the value measured toward the LMC.
At present, the dark halo MF is completely unknown but both the Hubble Deep Field
observations and the narrow-range of the observed time-distribution of the microlensing
events toward the LMC strongly suggest an IMF different from a Salpeter IMF below $\sim 1\,\msol$ (Chabrier 1999). Assuming a homogeneous distribution, both constraints yield a mass-density for LMS in the dark halo about 2 orders of magnitude smaller than the afore-mentioned spheroid one (Chabrier \& M\'era 1997). This means that essentially no stars formed below about a solar mass in the dark halo.

\subsection{Brown dwarf mass function}

Even though the possibility that BDs could make up for the Galactic missing mass is now clearly excluded, a proper census of the number of BDs has significant implications for our understanding of how stars
and planets form. The determination of the BD MF is a complicated task. By definition, BDs never reach
thermal equilibrium and most of the BDs formed at the early stages of the Galaxy will
have dimmed to very low-luminosities ($L\propto 1/t$).
Thus observations will be biased toward young and massive BDs.
This age-undetermination is circumvented when studying the BD MF in clusters because objects in this case are likely to be coeval. The Pleiades cluster has been extensively surveyed and several BDs have been identified down to $\sim 0.04\,\msol$ (Mart\'\i n et al
1998, Bouvier et al 1998, Hambly et al 1999). A single power-law function from $\sim 0.4$ to $0.04\,\msol$ seems to adequately reproduce the observations (not corrected for binaries) with some remaining uncertainties in the exponent:
$\alpha \sim 0.6-1.0$. However, stellar objects in very young clusters ($\simle 10^6$ yr) might still be accreting whereas older clusters may have already experienced significant dynamical evolution, mass segregation in the core and/or evaporation in the outer regions (see e.g. Raboud \& Mermilliod 1998).
In this case the present-day mass function does not reflect the initial
mass function.
For these
reasons, it is probably premature to claim a robust determination of the MF in young clusters.

The DENIS (Delfosse et al 1999b) and 2MASS (Kirkpatrick et al 1999b; Burgasser et al 1999) surveys, which have covered an area of several hundred square degrees and are complete to $K=13.5$ and 14.5, respectively, have revealed about 20 field L-dwarfs and 4 field methane-BDs. This yields an L-dwarf number-density $n_L\approx 0.03\pm 0.01$ sq deg$^{-1}$ for $K<14.5$ and a methane-BD number-density $n_{CH_4}\approx 0.002$ sq deg$^{-1}$ for $J<16$. Although these numbers correspond to small statistics and should be considered with caution, they provide the first observational constraints on the substellar MF in the Galactic disk and are consistent with a slowly rising BD MF
with $\alpha\simeq 1$-2 (Reid et al 1999; Chabrier in preparation), assuming a constant formation rate.

Independent, complementary information on the stellar and substellar MF comes from microlensing
observations. Indeed,
the time distribution of the events provides a (model-dependent) determination of
the mass-distribution and thus of the minimum mass of the dark objects:
$dN_{ev}/dt_e=E\times \epsilon (t_e)\times d\Gamma/dt_e \propto P(m)/\sqrt m$, where
$E$ is the observed exposure, i.e. the number of star$\times$years, $\epsilon$ is the
experimental efficiency, $\Gamma$ is the event rate and $P(m)$ is the mass probability
distribution.
The analysis of the published 40 MACHO (Alcock et al 1997) + 9 OGLE (Udalsky et al 1994) events toward the bulge is consistent with a rising MF at the bottom
of the MS whereas a decreasing MF
below 0.2 $\msol$ seems to be excluded at a high ($>90\%$) confidence level (Han \& Gould 1996, M\'era et al 1998).
Although the time distribution might be affected by various biases
(e.g. blending) and robust conclusions must wait for larger statistics,
the present results suggest
that, in order to explain
both star counts and the microlensing experiments, a substantial number of
BDs must be present in the Galactic disk. Extrapolation of the stellar MF determined in \S5.1 into the
BD domain down to 0.01 $\msol$ yields for the Galactic disk a BD {\it number}-density comparable to
the stellar one, $n_{BD}\approx 0.1\,pc^{-3}\simeq n_\star$, and a mass-density $\rho_{BD}\approx 3.0\times 10^{-3}\,
\mvol$, i.e. $\Sigma_{BD}\approx 2\,\msurf$.

For the spheroid, extrapolation of the previously-determined stellar MF yields $n_{BD}\simle 10^{-4}\,pc^{-3}$, $\rho_{BD}<  10^{-5}\,\mvol$, less than 0.1\% of the required
dynamical density. For the dark halo the density
is about 2 orders of magnitude smaller, as mentioned before.

It is obviously premature to try to infer the mass distribution of exoplanets. This will
first require a clear theoretical and observational distinction between planets and brown dwarfs. However, an interesting
preliminary result comes from the observed mass distribution of the companions of $G$ and $K$ stars. As shown in figure 4 of Mayor et al (1998), there is a
strong discontinuity in the mass distribution at $m_2/{\rm sin}\, i\approx 5\,\mj$, with a clear
peak below this limit. This suggests that planet formation in a
protoplanetary disk is a much more efficient mechanism than BD formation as star companions, at least around $G$ and $K$ stars (see e.g. Marcy \& Butler 1998). It also suggests that the MF for BD stellar companions differs from the MF in the field. Ongoing observations around M-stars will tell us whether such a mass distribution still holds around low-mass stars.

\section{Conclusions}

This review has summarized the significant progress achieved within the past few years in the theory of cool and dense objects at the bottom of and beyond the main sequence: low-mass stars, brown dwarfs and gaseous planets. The successful confrontation of the theory with the numerous detections of low-mass stellar and sub-stellar objects allows a better understanding of their structural and thermal properties, and allows reliable predictions about their evolution. This in turn brings confidence in the predicted characteristic properties of these objects, a major issue in terms of search strategies for future surveys.
Important problems remain to be solved to improve the theory. A non-exhaustive list includes for example: (i) a better determination of the EOS in the pressure-ionization region, with the possibility of a first-order phase transition, (ii) the study of phase separation of elements in SSO interiors, (iii) an improved treatment of convection in optically-thin regions, (iv) a precise description of the dynamics of grain formation and sedimentation in SSO atmospheres, (v) the derivation of an accurate mass-$\te$-age scale for SSOs and young objects, (vi) a correct understanding of magnetic-field generation and dissipation in active LMS and BDs.

The increasing number of observed LMS and SSOs, together with the derivation of accurate
models, eventually will allow a robust determination of the stellar and substellar mass
functions, of the minimum mass for the formation of star-like objects, and thus of the exact density of these objects in the Galaxy. As we discussed in \S 5,
present determinations in various Galactic regions point to a slowly rising MF near and
below the H-burning limit, with a BD number-density comparable to the stellar one, and a MF truncated below $\sim 1\,\msol$ in the dark halo. A more precise determination must await confirmation
from future observations. At last, the amazingly rapid pace of exoplanet discoveries
should yield the determination of the planetary MF and maximum mass, and eventually the direct detection of such objects, a future formidable test for the theory.
\bigskip

{\bf Acnowledments}: This review has benefited from various discussions with our colleagues F Allard, PH Hauschildt, D Alexander, K Lodders, A Burrows, J
Lunine, B Fegley, T Forveille, X Delfosse, J Bouvier, P Kroupa, A Nordlund, T Guillot, D Saumon, U Kolb. Our profound gratitude goes to these individuals, who helped by improving the original manuscript. We are also very indebted to P Kroupa, N Reid, HC Harris and C Dahn for providing various data prior to publication.

\eject

\centerline{\it Literature Cited}
\bigskip

\bib Alcock C, Allsman RA, Alves D, Axelrod TS, Bennett DP, et al. 1997. \apj. 479:119
\bib Alexander DR, Ferguson JW. 1994 \apj. 437:879
\bib Allard F. 1990, PhD thesis. Univ. Heidelberg, Germany
\bib Allard, F. 1999, {\it Very Low-Mass Stars and Brown Dwarfs
              in Stellar Clusters and Associations. Proc. Euroconference, La Palma, 1998}
\bib Allard F, Alexander DR, Tamanai A, Hauschildt P. 1998. 
{\it Brown dwarfs and extrasolar planets.  Proc. ASP Conf. Series}. 134:438
\bib Allard F, Hauschildt PH. 1995. \apj. 445:433
\bib Allard F, Hauschildt PH, Alexander DR, Starrfield S. 1997. \araa. 35:137
\bib Allard F, Hauschildt PH, Alexander DR, Tamanai A, 
Ferguson J. 2000. \apj. in preparation
\bib Allard F, Hauschildt PH, Baraffe I, Chabrier G. 1996. \apj. 465:L123
\bib Auman J. 1969. {\it Low Luminosity Star}. New York: Gordon and Breach
\bib Baraffe I, Chabrier G. 1996. \apj. 461:L51
\bib Baraffe I, Chabrier G, Allard F, Hauschildt PH. 1995. \apj. 446:L35
\bib Baraffe I, Chabrier G, Allard F, Hauschildt PH. 1997. \aap. 327:1054
\bib Baraffe I, Chabrier G, Allard F, Hauschildt PH. 1998. \aap. 337:403
\bib Basri G, Marcy GW, Graham JR. 1996. \apj. 458:600
\bib Basri G, Mohanty S, Allard F, Hauschildt PH, Delfosse X, et al. 2000. \apj.
 in press
\bib B\'ejar VJS, Zapatero Osorio MR, Rebolo R. 1999. \apj. 521:671
\bib Beuermann K, Baraffe I, Kolb U, Weichhold M. 1998. \aap. 339:518
\bib Bildsten L, Brown EF, Matzner CD, Ushomirsky G. 1997. \apj. 482:442
\bib Borysow A, Jorgensen UG, Zheng C. 1997. \aap. 324:185
\bib Bouvier J, Stauffer JR, Mart\'\i n EL, Barrado y Navascues B, Wallace B, B\'ejar VJS. 1998. \aap. 336:490
\bib Brett C. 1995. \aap. 295:736
\bib Brett C, Plez B. 1993. {\it Astron. Soc. Australia Proc.} 10:250
\bib Burgasser AJ, Kirkpatrick JD, Brown NE, Reid IN, et al. 1999. \apj. 522:65
\bib Burrows A, Hubbard WB, Lunine JI. 1989. \apj. 345:939
\bib Burrows A, Hubbard WB, Saumon D, Lunine JI. 1993. \apj. 406:158 
\bib Burrows A, Hubbard WB,  Lunine JI. 1994. {\it Cool Stars, Stellar Systems
and the Sun. ASP}. 64:528
\bib Burrows A, Liebert J. 1993. {\it Rev. Mod. Phys.} 65:301
\bib Burrows A, Marley M, Hubbard WB, Lunine JI, Guillot T, et al. 1997. \apj.
491:856
\bib Burrows A, Marley M, Sharp CM. 2000. \apj. 531:438
\bib Burrows A, Sharp CM. 1999. \apj. 512:843 
\bib Canuto VM, Mazzitelli I. 1991. \apj. 370:295
\bib Canuto VM, Goldman I, Mazzitelli I. 1996. \apj.  473:550
\bib Caughlan GR, Fowler WA. 1988. {\it Atom. Data and Nucl. Data Tables} 40:283
\bib Chabrier G. 1990. {\it J. de Physique} 51:1607
\bib Chabrier G. 1999. \apj. 513:L103
\bib Chabrier G, Baraffe I. 1997. \aap. 327:1039
\bib Chabrier G, Baraffe I, Allard F, Hauschildt PH. 2000. \apj. in press
\bib Chabrier G, Baraffe I, Plez B. 1996. \apj. 459:L91
\bib Chabrier G, M\'era D. 1997. \aap. 328:83
\bib Clausen JV, Helt BE, Olsen EH. 1999. 
{\it Theory and Tests of Convection
 in Stellar Structure. it  
   ASP Conf. Series} 173, American Society of Physics 
\bib Clayton DD. 1968. {\it Principles of Stellar Evolution and Nucleosynthesis}, Chicago Press
\bib Clemens JC, Reid IN, Gizis JE, O'Brien MS. 1998. \apj. 496:352
\bib Collins GW, Da Silva LB, Celliers P, Gold DM, Foord ME, et al. 1998. {\it Science} 281:1178
\bib Copeland H, Jensen JO, Jorgensen HE. 1970. \aap. 5:12
\bib Cox JP, Guili RT. 1968. {\it Principles of Stellar Structure}.
New York: Gordon and Breach
\bib Clemens JC, Reid IN, Gizis JE, O'Brien MS. 1998. ApJ 496:392
\bib Dahn CC, Guetter HH, Harris HC, Henden AA, CB Luginbuhl, et al. 1999. {\it
The Evolution of Galaxies on Cosmological Timescales. ASP Conf. Series.}
\bib Dahn CC,  Liebert J, Harris HC, Guetter HH. 1995. 
{\it The Bottom of the Main Sequence and Beyond. ESO Astrophys. Symp.} Berlin/Heidelberg: Springer Verlag, p.239
\bib Dahn CC, Liebert J,  Harrington RS. 1986. \aj. 91:621
\bib Da Silva, et al. 1997. {\it Phys. Rev. Lett.} 78:483
\bib D'Antona F, Mazzitelli I. 1985. \apj. 296:502
\bib Delfosse X, Forveille T, Beuzit JL, Udry S, Mayor M, Perrier C. 1999a.
\aap. 344:897
\bib Delfosse X, Forveille T, Perrier C, Mayor M. 1998a. \aap. 331:581
\bib Delfosse X, Forveille T, Tinney CG, Epchtein N. 1998b. See Allard et al. 1998, p.67
\bib Delfosse X, Tinney CG, Forveille T, Epchtein N, Bertin E, et al. 1997. \aap 327:L25
\bib Delfosse X, Tinney CG, Forveille T, Epchtein N, Borsenberger J,
et al. 1999b. \aaps. 135:41
\bib Delfosse X, Forveille T, Udry S, Beuzit JL, Mayor M, Perrier C. 1999c.
 \aap. {\it Lett.} 350:L39
\bib Demarque P, Guenther DB, Kim Y-C. 1999. \apj. 517:510
\bib Dorman B, Nelson LA, Chau WY. 1989. \apj. 342:1003
\bib Durney B, De Young DS, Roxburgh IW. 1993. {\it So. Ph.} 145:207
\bib Fegley B, Lodders K. 1994. {\it Icarus} 110:117
\bib Fegley B, Lodders K. 1996. \apj. 472:L37
\bib Fontaine G, Graboske HC, VanHorn HM. 1977. \apjs. 35:293
\bib Forveille T, Beuzit JL, Delfosse X, Segransan D, Beck F, et al. 1999.
 \aap. 351:619
\bib Goldman B, Delfosse X, Forveille T, Afonso C, Alard C, et al. 1999. \aap.
351:L5
\bib Gould A, Flynn C, Bahcall JN. 1998. \apj. 503:798
\bib Graff D, Freese K. 1996. \apj. 456:49
\bib Griffith CA, Yelle RV, Marley MS. 1998. {\it  Science} 282:2063
\bib Grossman L. 1972. {\it Geochimica et Cosmochimica Acta} 36:597
\bib Grossman, A.S., Hays, D., \& Graboske, H.C. 1974, \aap, 30, 95
\bib Guillot T. 1995. {\it Science} 269:1697
\bib Guillot T. 1999. {\it Science} 286:72
\bib Guillot T, Burrows A, Hubbard WB, Lunine JI, Saumon D. 1996. \apj. 459:L35
\bib Hambly NC, Hodgkin ST, Cossburn MR, Jameson RF. 1999. \mnras. 303:835
\bib Han C, Gould A. 1996. \apj. 467:540
\bib Hauschildt PH, Allard F, Baron E. 1999. \apj. 512:377
\bib Hawley SL, Gizis JE, Reid IN. 1996. \aj. 112:2799
\bib Henry TJ, McCarthy DW. 1993. \aj. 106:773
\bib Henry TJ, Ianna PA, Kirkpatrick D, Jahreiss H. 1997. \aj. 114:388
\bib Hubbard WB. 1977. {\it Icarus} 30:305
\bib Hubbard WB. 1994. The equation of state in astrophysics.
{\it Proc. IAU Coll., Saint-Malo,1993,} 147:443.
Cambridge: Cambridge University Press
\bib H\"ummer DG, Mihalas D. 1988. \apj. 331:794
\bib Iglesias CA,  Rogers FJ. 1996. \apj. 464:943
\bib Jones HRA, Tsuji T. 1997. \apj  480:L39
\bib Kirkpatrick JD, Allard F, Bida T, Zuckerman B, Becklin EE,
et al. 1999b. \apj. 519:834
\bib Kirkpatrick JD, Henry TJ, McCarthy DW. 1991. \apjs. 77:417
\bib Kirkpatrick JD,  McCarthy DW. 1994. {\it AJ.} 107:333
\bib Kirkpatrick JD, Reid IN, Liebert J, Cutri RM, Nelson B,
et al. 1999a. \apj. 519:802
\bib King AR. 1988.  {\it QJRAS}. 29:1
\bib Kolb U, Baraffe I. 1999a. \mnras. 309:1034
\bib Kolb U, Baraffe I. 1999b. {\it Annapolis Workshop on Magnetic Cataclysmic
Variables. ASP Conf. Series. } 197:273
\bib Kolb U, King AR, Ritter H. 1998. \mnras. 298:L29
\bib Kroupa P. 1995. \apj. 453:350 
\bib Kroupa P, Tout CA, Gilmore G. 1990. \mnras. 244:76
\bib Kroupa P, Tout CA, Gilmore G. 1993. \mnras. 262:545
\bib Kumar SS. 1963. \apj. 137:1121
\bib Leggett S. 1992. \apjs. 82:351
\bib Leggett S, Allard F, Berriman G, Dahn CC, Hauschildt PH. 1996. \apjs. 104:117
\bib Leggett S, Allard F, Hauschildt PH. 1998. \apj. 509:836
\bib Lenzuni P, Chernoff DF, Salpeter EE. 1991. \apjs 76:759
\bib Leung KC, Schneider D.  1978. {\it AJ}. 83:618
\bib Linsky J. 1969. \apj. 156:989
\bib Lodders K. 1999. \apj. 519:793
\bib Ludwig HG, Freytag B, Steffen M. 1999. \aap. 346:111
\bib Luhman KL. 1999. \apj. 525:466
\bib Lunine JI, Hubbard WB, Burrows A, Wang YP, Garlow K. 1989. \apj. 338:314
\bib Lunine JI, Hubbard WB, Marley MS. 1986. \apj. 310:238
\bib Magni G,  Mazzitelli I. 1979. \aap. 72:134
\bib Marcy GW, Butler RP. 1998. \araa. 36:57
\bib Marley MS, Gelino C, Stephens D, Lunine JI, Freedman R. 1999. \apj. 513:879
\bib Marley MS, Hubbard WB. 1988. {\it Icarus} 73:536
\bib Marley MS, Saumon D, Guillot T, Freedman RS, Hubbard WB, et al. 1996. {\it Science} 272:1919
\bib Mart\'\i n E. 1999. See Allard 1999.
\bib Mart\'\i n EL,  Basri G, Delfosse X, Forveille T. 1997. \aap. 327:L29
\bib Mart\'\i n EL,  Basri G, Zapatero-Osorio MR, Rebolo R, Lopez RJ.
 1998. \apj. 507:L41
\bib Maxted PFL, Marsh TR, Moran C, Dhillon VS,
     Hilditch RW. 1998. \mnras. 300:1225
\bib Mayor M,  Queloz D. 1995. {\it Nature} 378:355
\bib Mayor M, Queloz D, Udry S. 1998. See Allard et al 1998, p.140
\bib Metcalfe TS, Mathieu RT, Latham DW, Torres G. 1996.
     \apj. 456:356
\bib M\'era D, Chabrier G, Schaeffer R. 1998. \aap. 330:937
\bib Meyer F, Meyer-Hofmeister E. 1999. \aap. 341:L23
\bib Miller G, Scalo J. 1978. {\it PASP}. 90:506
\bib Monet D, Dahn CC, Vrba FJ, Harris HC, Pier JR, et al. 1992. {\it AJ}. 103:638
\bib Nelson LA, Rappaport S, Joss PC. 1986. \apj. 311:226
\bib Nelson LA, Rappaport S, Chiang E. 1993b. \apj. 413:364 
\bib Nelson LA, Rappaport S, Joss PC. 1993a. \apj. 404:723
\bib Neuh\"auser R, Briceno C, Comeron F, Hearty T, Mart\'\i n EL, et al 1999. 
\aap. 343:883
\bib Neuh\"auser R, Comeron F. 1998, {\it Science} 282:83
\bib Noll KS, Geballe TR, Marley MS, 1997, \apj, 489:L87
\bib Nordlund \aa, Stein RF. 1999. See Clausen et al 1999, p.91 
\bib Oppenheimer BR, Kulkarni SR, Mathews K, Nakajima T. 1995. {\it Science} 270:1478
\bib Oppenheimer BR, Kulkarni SR, Mathews K, VanKerkwijk MH. 1998 \apj 502:932
\bib Parker EN. 1955. \apj. 122:293
\bib Parker EN. 1975. \apj. 198:205
\bib Pulone L, De Marchi G, Paresce F, Allard F. 1998. \apj. 492:L41
\bib Raboud D, Mermilliod JC. 1998. \aap. 333:897
\bib Randich S. 1999. See Allard 1999.
\bib Rappaport S, Verbunt F, Joss PC. 1983. \apj. 275:713
\bib Rebolo R, Mart\'\i n EL, Magazz\`u A. 1992. \apj. 389:L83
\bib Rebolo R, Zapatero Osorio MR, Mart\'\i n EL. 1995. {\it Nature} 377:129
\bib Reid IN. 1999. \araa. 37:191
\bib Reid IN, Gizis JE. 1997. \aj. 113:2246
\bib Reid I.N, Hawley SL, Gizis JE. 1995. \aj. 110:1838
\bib Reid IN, Kirkpatrick JD, Liebert J, Burrows A, Gizis JE, et al.
 1999. \apj. 521:613
\bib Rossow WB. 1978. {\it Icarus} 36:1
\bib Ruiz MT, Leggett SK, Allard F. 1997. \apj. 491:L107
\bib Salpeter EE. 1954. {\it Austr. J. Phys.} 7:373
\bib Salpeter EE. 1961. \apj. 134:669
\bib Saumon D. 1994. See Hubbard 1994, p.306
\bib Saumon D, Bergeron P, Lunine JI, Hubbard WB, Burrows A. 1994. \apj. 424:333
\bib Saumon D, Chabrier G. 1991. {\it Phys. Rev. A}. 44:5122
\bib Saumon D, Chabrier G. 1992. {\it Phys. Rev. A}. 46:2084
\bib Saumon D, Chabrier G, VanHorn HM. 1995. \apjs. 99:713
\bib Saumon D, Hubbard WB, Burrows A, Guillot T, Lunine JI, Chabrier G. 1996.
\apj. 460:993
\bib Schatzman E. 1948. {\it J. Phys. Rad.} 9:46
\bib Schultz AB, Allard F, Clampin M, McGrath M, Bruhweiler FC, et al. 1998.
\apj. 492:L181
\bib Seager S, Sasselov DD. 1998. \apj. 502:L157
\bib Sharp CM,  Huebner WF. 1990. \apjs. 72:417
\bib Shu FH, Adams FC, Lizano S. 1987. \araa. 25:23
\bib Spiegel EA, Weiss NO. 1980. {\it Nature} 287:616
\bib Spiegel EA, Zahn J-P. 1992 \aap. 265:106
\bib Spruit H. 1994. {\it Cosmical Magnetism. NATO ASI Series C.} 422:33-44
\bib Spruit H, Nordlund \aa. 1990. \araa. 28:263
\bib Spruit H, Ritter H. 1983. \aap. 124:267
\bib Stauffer JR, Hartmann LW, Prosser CF, Randich S, Balachandran S, 
 et al. 1997. \apj. 479:776
\bib Stauffer JR, Schultz G,  Kirkpatrick JD. 1998. \apj. 499:L199
\bib Stein RF, Nordlund \aa. 2000. {\it Solar Phys. } 192:91 
\bib Stevenson DJ. 1979. {\it Geophys. Astrophys. Fluid Dyn.} 12:139
\bib Stevenson DJ. 1991. \araa. 29:163
\bib Taam RE, Spruit HC. 1989. \apj. 345:972
\bib Tarter J. 1975. PhD thesis. Univ. Calif., Berkeley
\bib Tinney CG, Delfosse X, Forveille T, Allard F. 1998. \aap. 338:1066
\bib Tinney CG, Reid IN. 1998. \mnras. 301:1031
\bib Toth RA. 1997. {\it  J. Mol. Spect.} 186:276
\bib Tsuji T, Ohnaka K, Aoki W. 1995. See Dahn et al 1995, p.45
\bib Tsuji T, Ohnaka K, Aoki W. 1996. \aap. 305:L1
\bib Tsuji T, Ohnaka K, Aoki W. 1999. \apj. 520:L119
\bib Udalsky A, Szymanski M, Stanek KZ, Kaluzny J, Kubiak M, et al. 1994. {\it Acta Astronomica} 44:165
\bib Ushomirsky G, Matzner CD, Brown EF, Bildsten L, Hilliard VG, 
Schroeder P. 1998. \apj. 253:266
\bib VandenBerg DA, Hartwick FDA, Dawson P, Alexander DR.
1983. \apj. 266:747
\bib White RJ, Ghez AM, Reid IN, Schultz G. 1999. \apj. 520:811
\bib Zapatero Osorio MR, Bejar VJS, Rebolo R, Mart\'\i n EL, Basri G. 1999. \apj.
524:L115
\bib Zapolsky HS, Salpeter EE. 1969. \apj. 158:809

\eject

\begin{table}
\caption{ZAMS age for solar metallicity LMS (after Baraffe et al (1998))}
\begin{tabular}{cccccccccc}
\hline\noalign{\smallskip}
$m/\msol$  & 0.075 & 0.08 & 0.09 & 0.1 & 0.2 & 0.4 & 0.6 & 0.8 & 1 \\
$t_{\rm ZAMS}$ (Gyr) & 3  & 1.7  & 1 & 0.7  & 0.33  & 0.20  & 0.12  & 0.06  & 0.04 \\
\hline
\end{tabular}
\label{table.zams}
\end{table}

\vfill\eject\clearpage

\begin{figure}
\centerline{\it Figure Legends}

\caption[]{Central temperature (in K), density (in g cm$^{-3}$), and degeneracy parameter along the LMS-SSO mass range for objects with $Z=\zsol$ at
5 Gyr (solid line) and 10$^8$ yr (dashed line), and with a metallicity $Z=10^{-2} \zsol$ at 5 Gyr (dotted-line).}
\label{chabF1.eps}
\end{figure}

\begin{figure}
\caption[]{Central temperature as a function of age for different
masses. $T_{\rm H}$, $T_{\rm Li}$ and $T_{\rm D}$ indicate the hydrogen, lithium and deuterium
 burning temperatures, respectively.}
\label{chabF2.eps}
\end{figure}

\begin{figure}
\caption[]{Mass-radius relationship for LMS and SSOs for 2 ages, $t=6\times 10^7$ yr (dash-dot) and 5$\times 10^9$ yr (solid) for $Z=Z_\odot$, and $t=5\times 10^9$ yr for $Z=10^{-2}\times Z_\odot$ (dash). The HBMM is 0.075 $\msol$ for $Z=Z_\odot$ and 0.083 $\msol$
for $Z=10^{-2}\times Z_\odot$.
Also indicated are the observationally-determined radii of various objects (see text)
and the
position of Jupiter radius (J). The bump on the $6\times 10^7$ yr isochrone
illustrates the initial D-burning phase.}
\label{chabF3.eps}
\end{figure}

\begin{figure}
\caption[]{Mass-$\te$ relationship for LMS and SSOs. Same legend as in 
Figure 3. The arrows indicate the onset of H$_2$ formation
near the photosphere. Note the log scale in the inset.}
\label{chabF4.eps}
\end{figure}

\begin{figure}
\caption[]{$T_{\rm c}$-$\rho_{\rm c}$ relationship (in cgs) for LMS (solid lines) and SSOs (dashed
lines) from 1 $\msol$ to 0.001 $\msol$ (masses in $\msol$ indicated on the curves).
Dotted lines represent 10$^6$, 10$^7$, 10$^8$, 10$^9$ and 
5 10$^9$ yr isochrones from bottom to top. The bumps on the 10$^6$-10$^8$ yr
isochrones at $\log \, T_{\rm c} \sim 5.4-5.8$ correspond to the initial
deuterium burning phase.}
\label{chabF5.eps}
\end{figure}

\begin{figure}
\caption[]{Effective temperature vs time (yr) for objects from 1 $\msol$ to 10$^{-3}$ $\msol$
(masses indicated in $\msol$). 
Solid lines: $Z=Z_\odot$, no dust opacity; dotted lines: $Z=Z_\odot$, dust opacity included, shown for 0.01, 0.04 and 0.07 $\msol$; 
dashed line:  $Z=10^{-2}\times Z_\odot$ (only for 0.3 $\msol$).}
\label{chabF6.eps}
\end{figure}

\begin{figure}
\caption[]{Theoretical H-R diagram for various
 masses (labeled in $\msol$). The weak dependence of radius on mass for SSOs yields the merging of the tracks for the lowest-mass objects. 
Dotted lines represent 10$^6$, 10$^7$, 10$^8$ and 
5 10$^9$ yr isochrones from right to left. The models are calculated with a 
mixing length $l \, = \, 1.9 \, H_{\rm P}$.
}
\label{chabF7.eps}
\end{figure}

\begin{figure}
\caption[]{$\log\, g$ (in cgs) vs $\te$ (in K) for LMS (solid curves) 
and SSOs (dashed
curves) from 1 $\msol$ to 0.001 $\msol$ (masses in $\msol$ indicated on the curves).
Dotted lines represent 10$^6$, 10$^7$, 10$^8$ and 
5 10$^9$ yr isochrones from bottom to top.
}
\label{chabF8.eps}
\end{figure}

\begin{figure}
\caption[]{Mass-M$_K$ relationship. Observationally determined masses are indicated by filled triangles (Henry \& McCarthy 1993) and circles
(Delfosse et al 1999c, Forveille et al 1999). Solid line: $\mz=0$, $t=5\times 10^9$ yr; dotted line: $\mz=0$, $t= 10^8$ yr; dash-line: $\mz=-0.5$, $t=10^{10}$ yr. Note the log-scale in the inset.} 
\label{chabF9.eps}
\end{figure}

\begin{figure}
\caption[]{Mass-Spectral type relationships. Observations are from: Kirkpatrick
 \& McCarthy
 (1994) for disk M-dwarfs (circles) and Clausen et al. (1999) for K- and G- dwarfs
(squares). Note that the latter objects are likely to be young. The Sun is included at  $Sp$ = G2. The theoretical models
are from Baraffe et al (1998). 
 Solid line: $\mz=0$, $t=5\times 10^9$ yr; dotted line: $\mz=0$, $t=10^8$ yr;
dot-dash-line: $\mz=0$, $t=3\times 10^{7}$ yr; 
dash-line: $\mz=-0.5$, $t=5\times 10^{9}$ yr.}
\label{chabF10.eps}
\end{figure}

\begin{figure}
\caption[]{Spectrum of a young EGP (T$_{\rm eff}=1000$K, defined as the effective temperature of the non-irradiated object) irradiated
by a G2V primary at $0.2, 0.1,$ and $0.05\,$AU distance (bottom to top). The plot
shows the incident spectrum (topmost curve), the spectrum emitted by
the irradiated planet (magenta curves with more flux and shallower
water and methane bands) and the non-irradiated spectrum of the EGP
(bottom full black curve) assuming a distance of 15 pc to Earth.
The dashed black line is the blackbody spectrum for T$_{\rm eff}=1000$K.
The EGP model includes dust opacities.
(courtesy of F. Allard and P.H. Hauschildt)}
\label{chabF11.eps}
\end{figure}

\begin{figure}
\caption[]{M$_V$ vs (V-I) diagram for different ages and metallicites:
DUSTY models (see text) for $\mz$=0 for 10$^8$ yr (dotted
line) and $5\times 10^9$ yr (solid line); grainless models for $\mz$ =-2 and t=10 Gyr (dashed
line)
The crosses, circles and squares correspond to 
observations from Dahn et al (1995), Monet et al (1992) and Dahn et al (1999), respectively. The indicated masses (in $\msol$) correspond to the 5 Gyr $\mz$=0 isochrone.
}
\label{chabF12.eps}
\end{figure}

\begin{figure}
\caption[]{M$_K$ vs (J-K) diagram for different ages and metallicites:
$\mz$=0 for 10$^8$ yr (dotted
lines) and $5\times 10^9$ yr (solid lines);  $\mz$ =-2, t=10 Gyr (dashed
line).
The red curves on the right correspond to the DUSTY models for $\mz$=0.
The blue curves on the left correspond to the COND models for $\mz$=0 (see text, \S 4.5.2).
Filled circles and triangles on the isochrones indicate masses either
in $\msol$ or $M_{\rm J}$ (1 $M_{\rm J}\approx 10^{-3}\,\msol$).
Small green circles:
   MS stars from Leggett (1992) and Leggett et al  (1996); 
black squares: Dahn et al (1999).
 Some identified BDs are also indicated (green diamonds).
}
\label{chabF13.eps}
\end{figure}

\begin{figure}
\caption[]{Deuterium-depletion curves as a function of age.
The dashed lines correspond to different SSO masses, while
the solid lines correspond to the 50\% and 99\% D-depletion
limit, respectively. The inset displays the corresponding $\te(t)$.}
\label{chabF14.eps}
\end{figure}

\begin{figure}
\caption[]{Luminosity function in the V-band from the 5-pc sample (dash-line) (Kroupa, unpublished), the 8-pc sample (solid-line) (Reid et al 1999) and from the HST (dot-line) (Gould et al 1998)}
\label{chabF15.eps}
\end{figure}

\begin{figure}
\caption[]{Mass function derived from the 5-pc LF (squares, last bin out of the figure) and the 8-pc LF (circles). Dash-line: $dn/dm=0.024\,m^{-1.2}$ (see text)}
\label{chabF16.eps}
\end{figure}

\vfill\eject\clearpage

\begin{figure}
\begin{center}
\epsfxsize=180mm
\epsfysize=200mm
\epsfbox{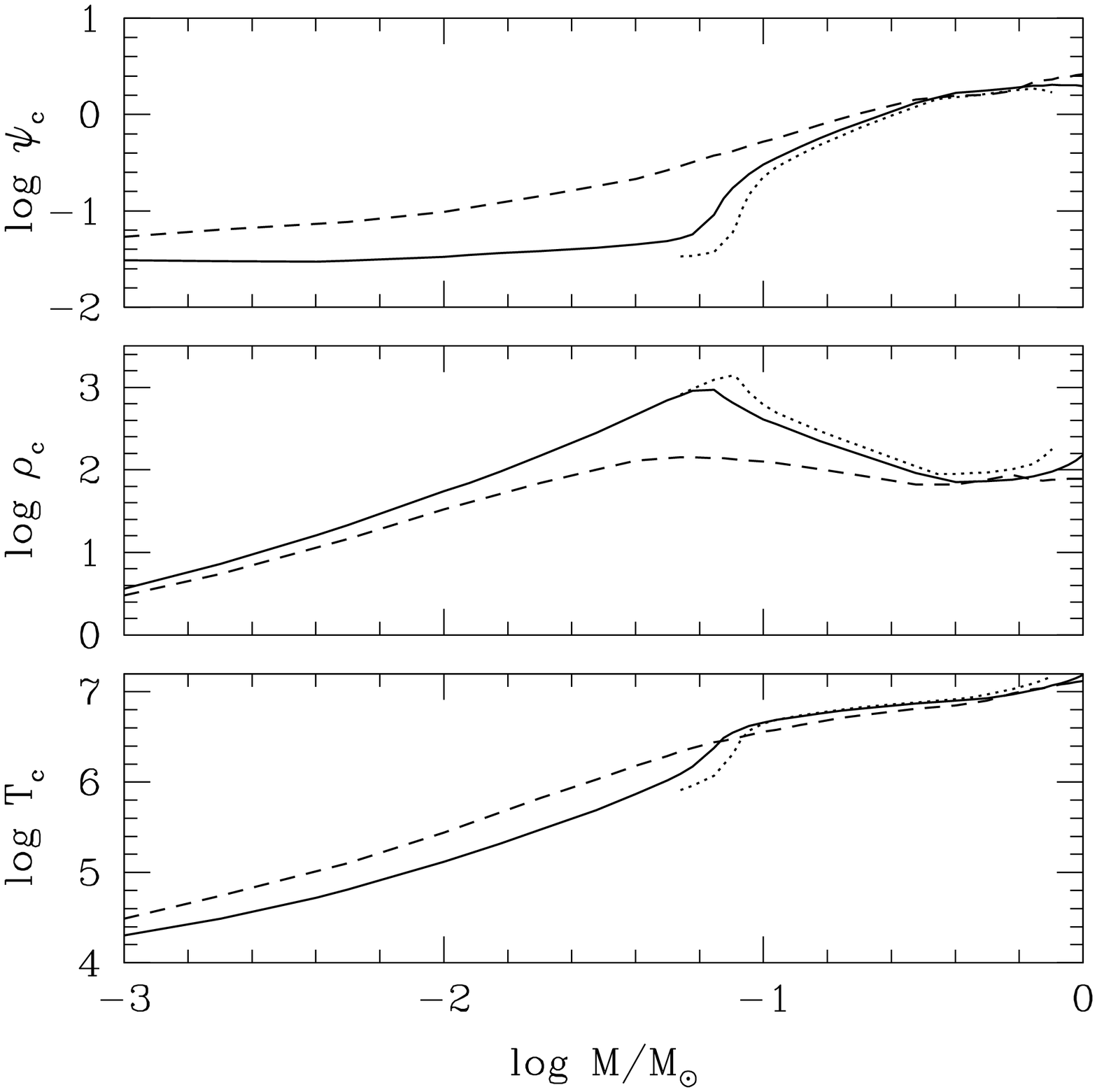}
\end{center}
\end{figure}

\vfill\eject

\begin{figure}
\begin{center}
\epsfxsize=180mm
\epsfysize=200mm
\epsfbox{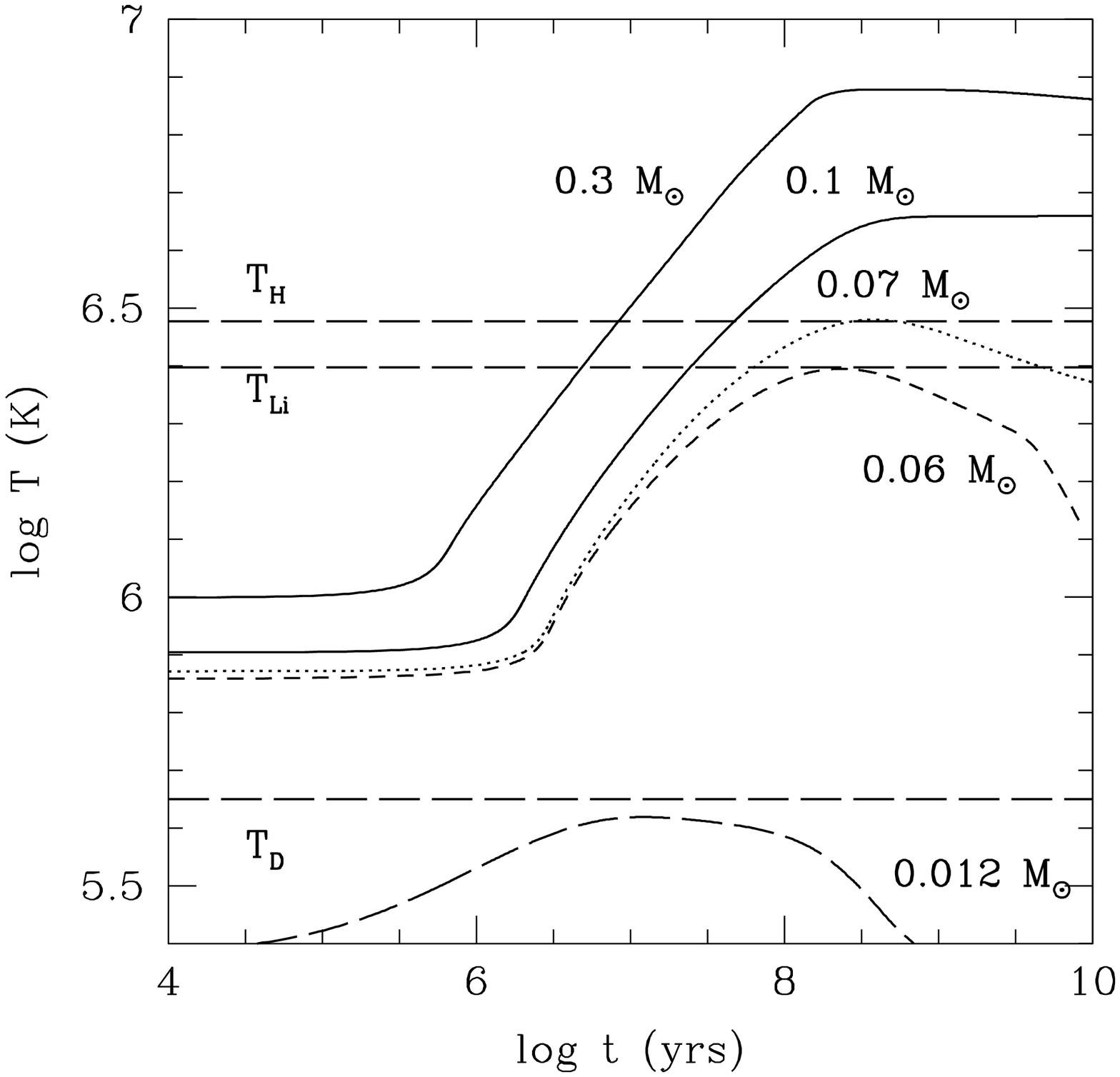}
\end{center}
\end{figure}

\vfill\eject

\begin{figure}
\begin{center}
\epsfxsize=180mm
\epsfysize=200mm
\epsfbox{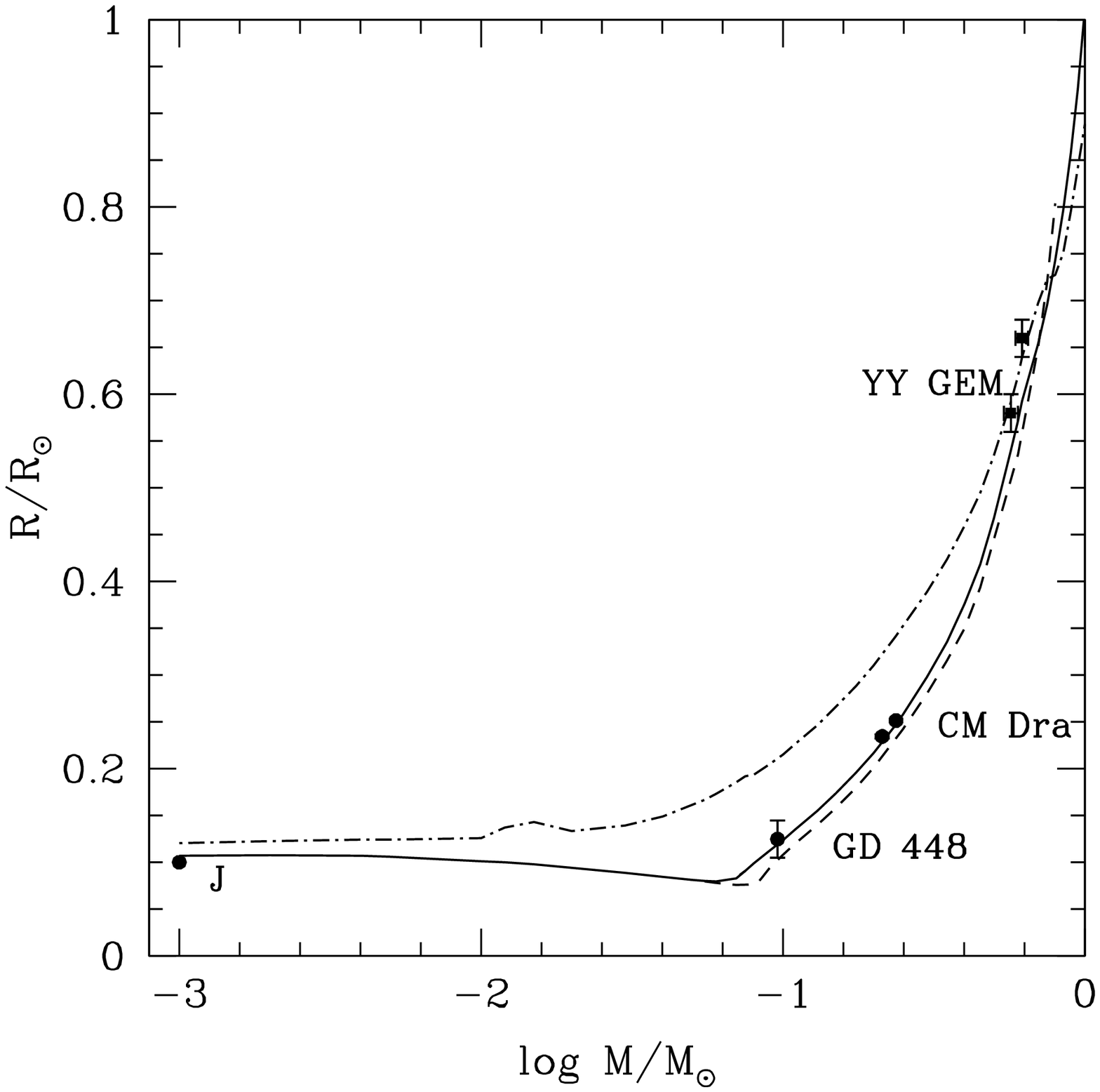}
\end{center}
\end{figure}

\vfill\eject

\begin{figure}
\begin{center}
\epsfxsize=180mm
\epsfysize=200mm
\epsfbox{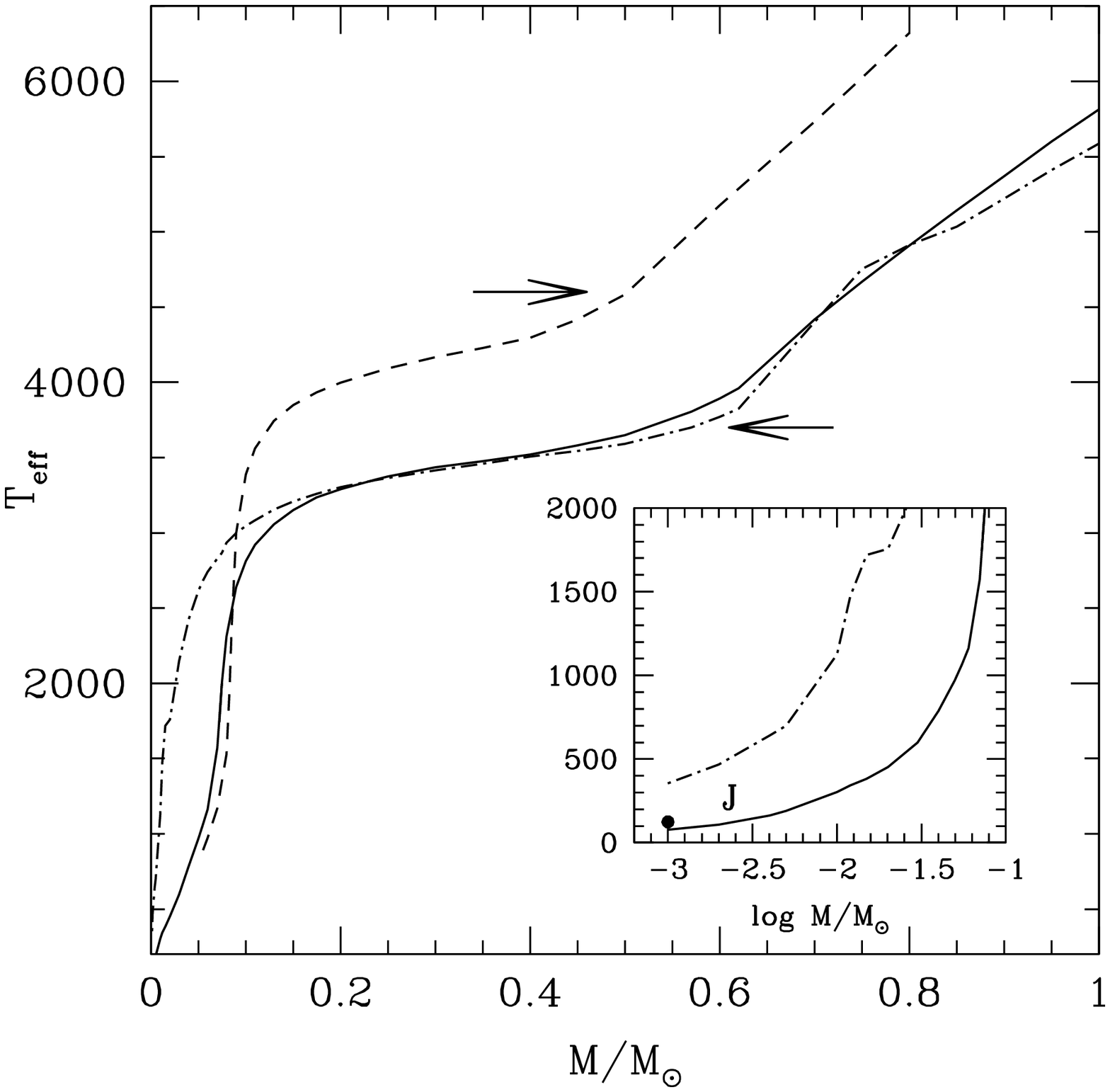}
\end{center}
\end{figure}

\vfill\eject

\begin{figure}
\begin{center}
\epsfxsize=180mm
\epsfysize=200mm
\epsfbox{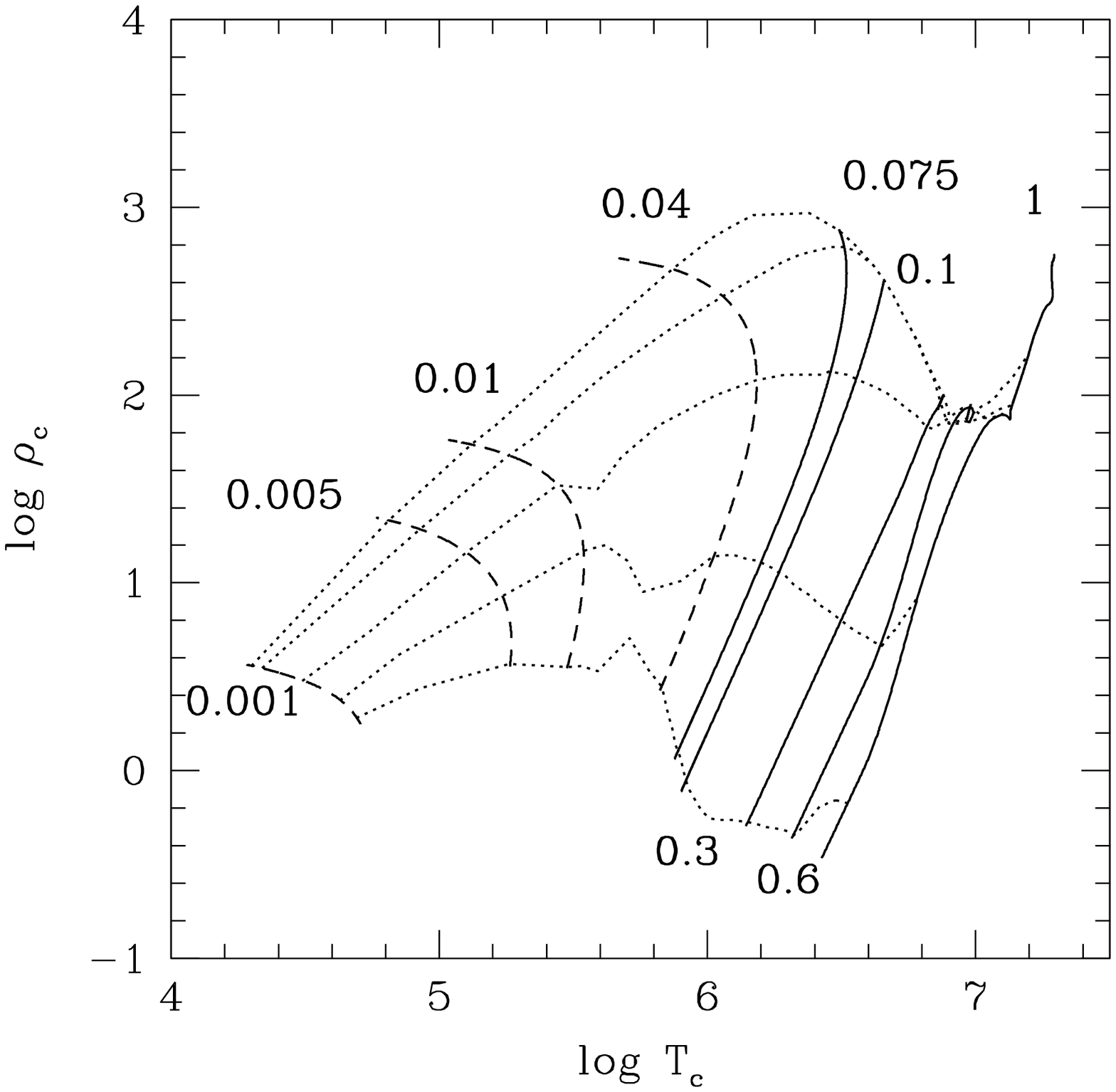}
\end{center}
\end{figure}

\vfill\eject

\begin{figure}
\begin{center}
\epsfxsize=180mm
\epsfysize=200mm
\epsfbox{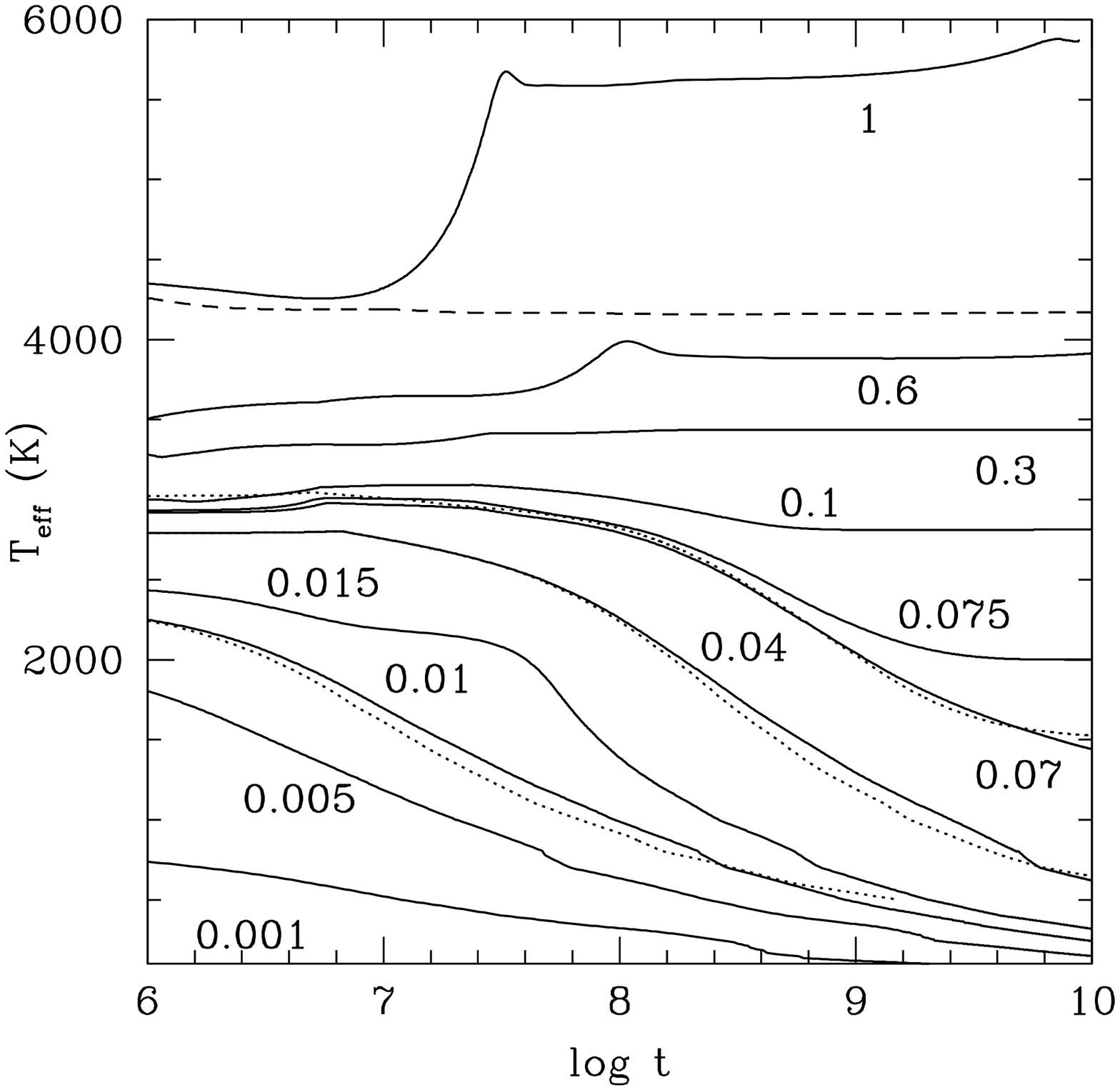}
\end{center}
\end{figure}

\vfill\eject

\begin{figure}
\begin{center}
\epsfxsize=180mm
\epsfysize=200mm
\epsfbox{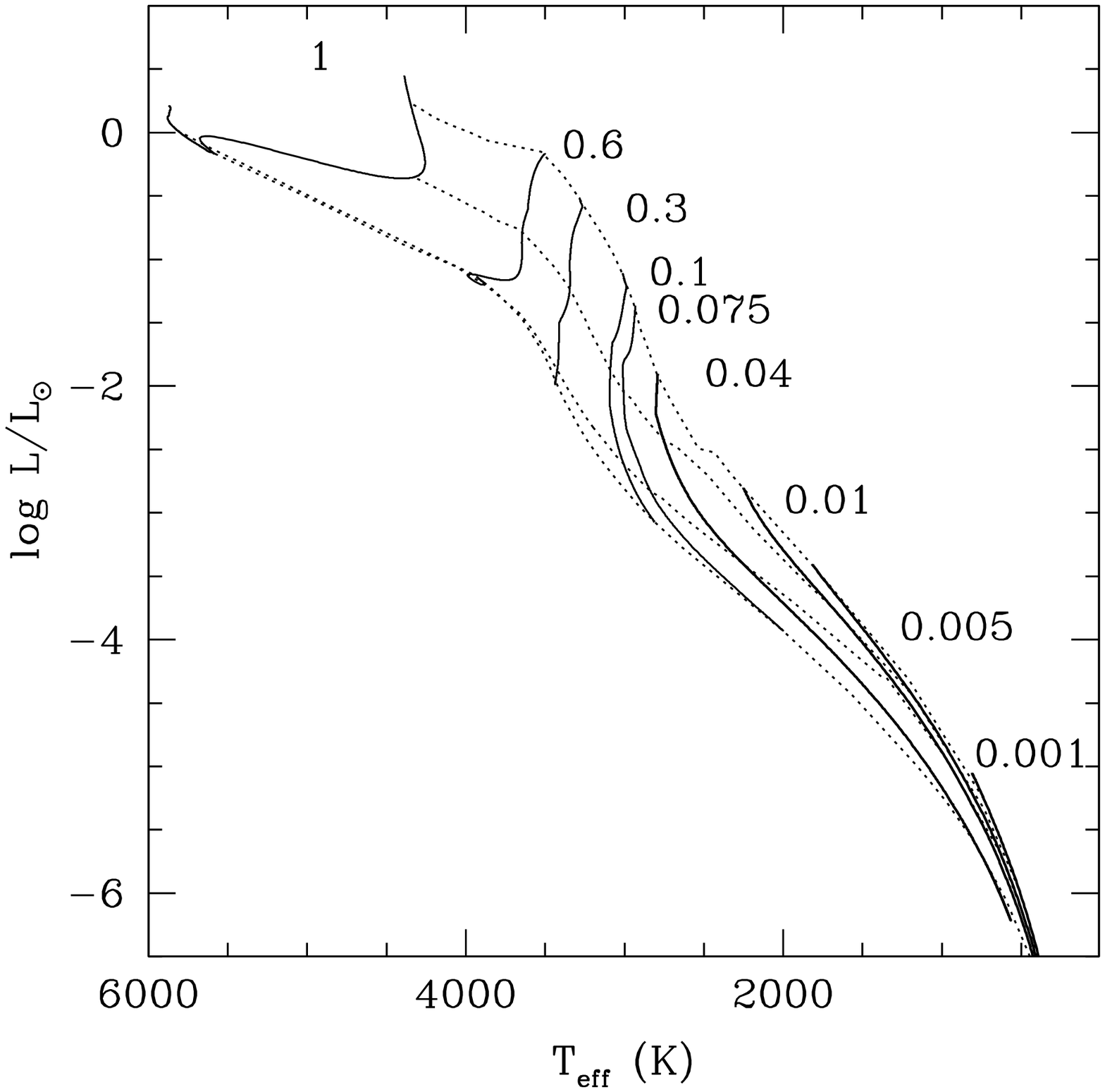}
\end{center}
\end{figure}

\vfill\eject

\begin{figure}
\begin{center}
\epsfxsize=180mm
\epsfysize=200mm
\epsfbox{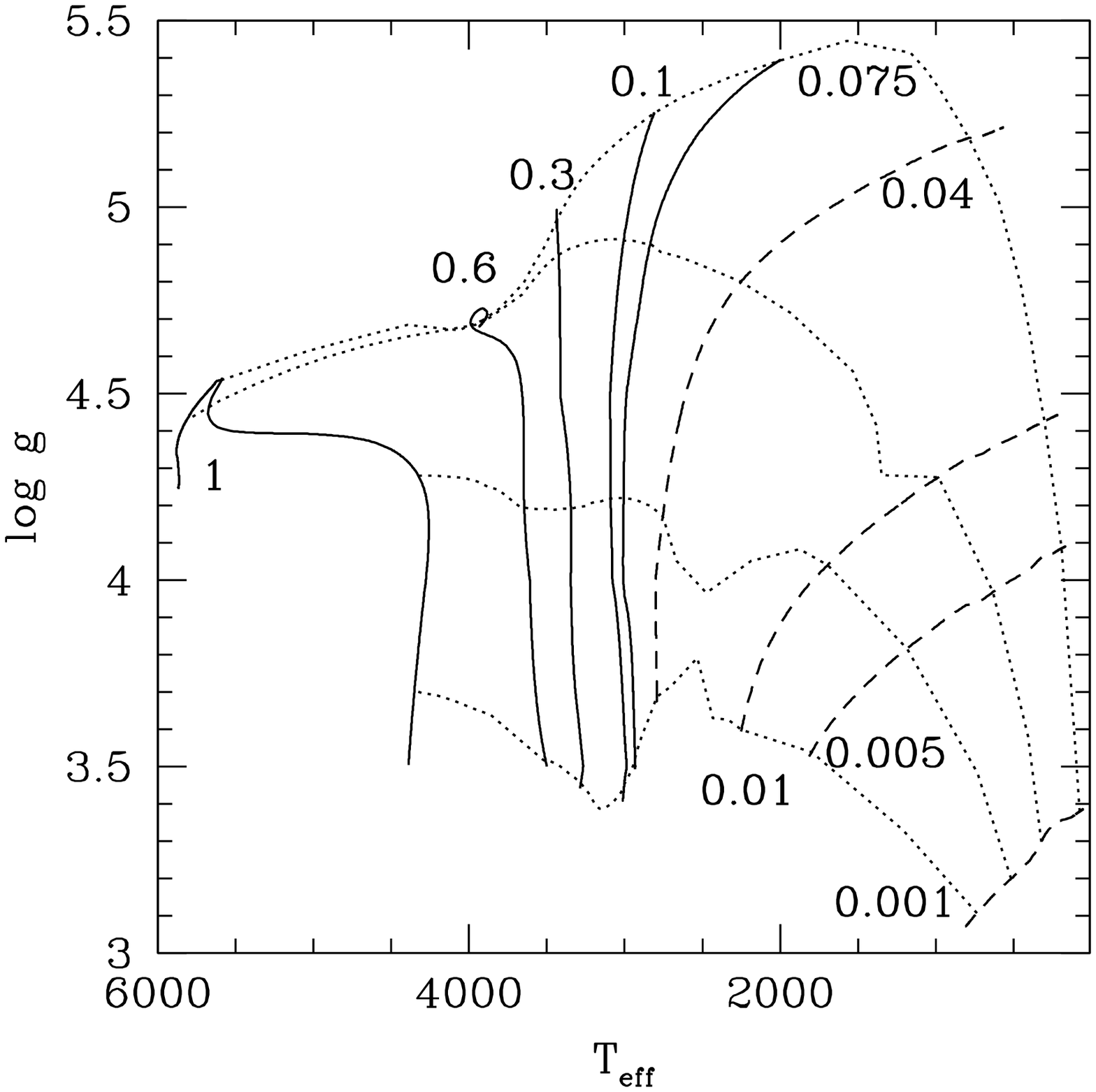}
\end{center}
\end{figure}

\vfill\eject

\begin{figure}
\begin{center}
\epsfxsize=180mm
\epsfysize=200mm
\epsfbox{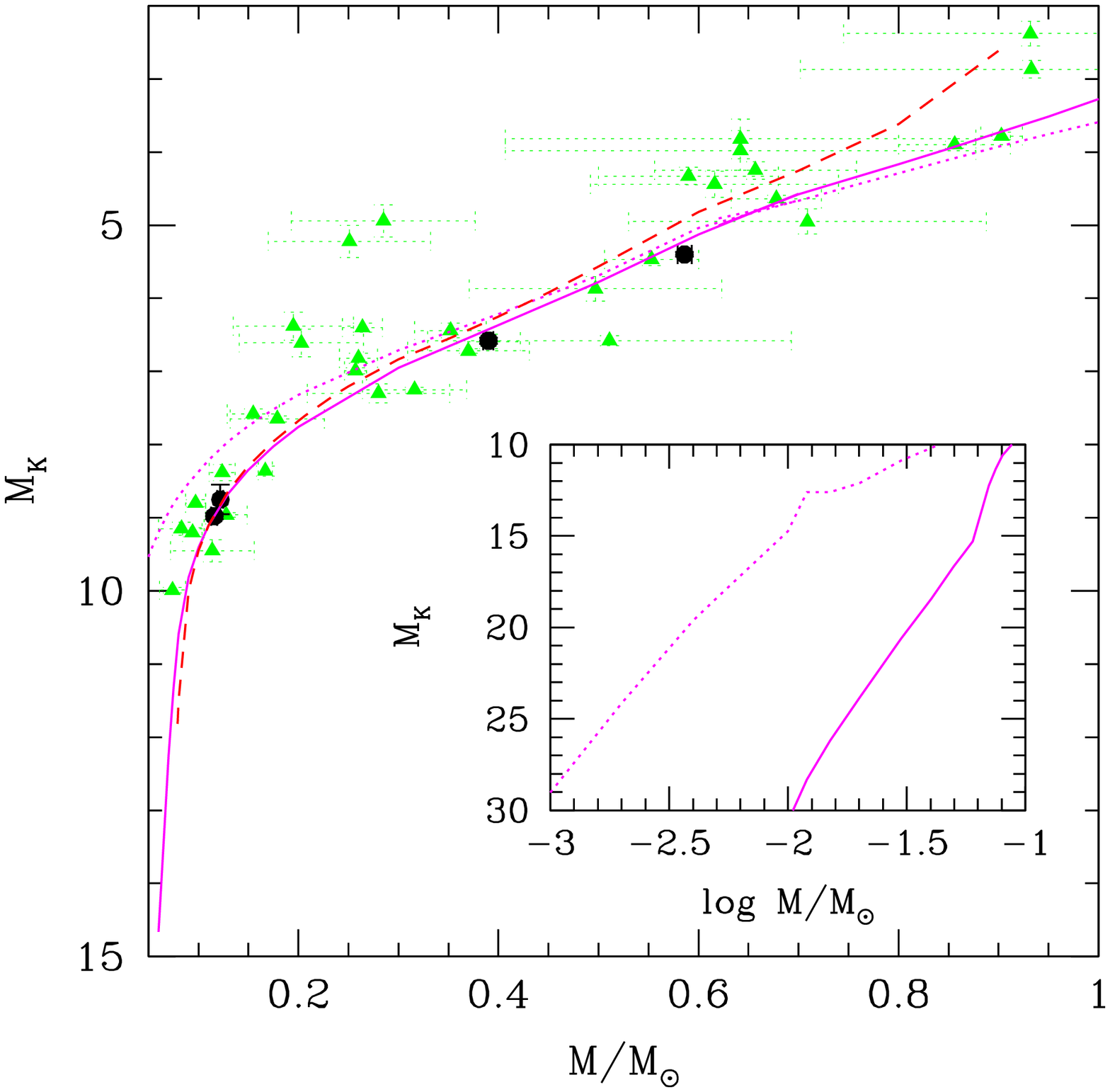}
\end{center}
\end{figure}

\vfill\eject

\begin{figure}
\begin{center}
\epsfxsize=180mm
\epsfysize=200mm
\epsfbox{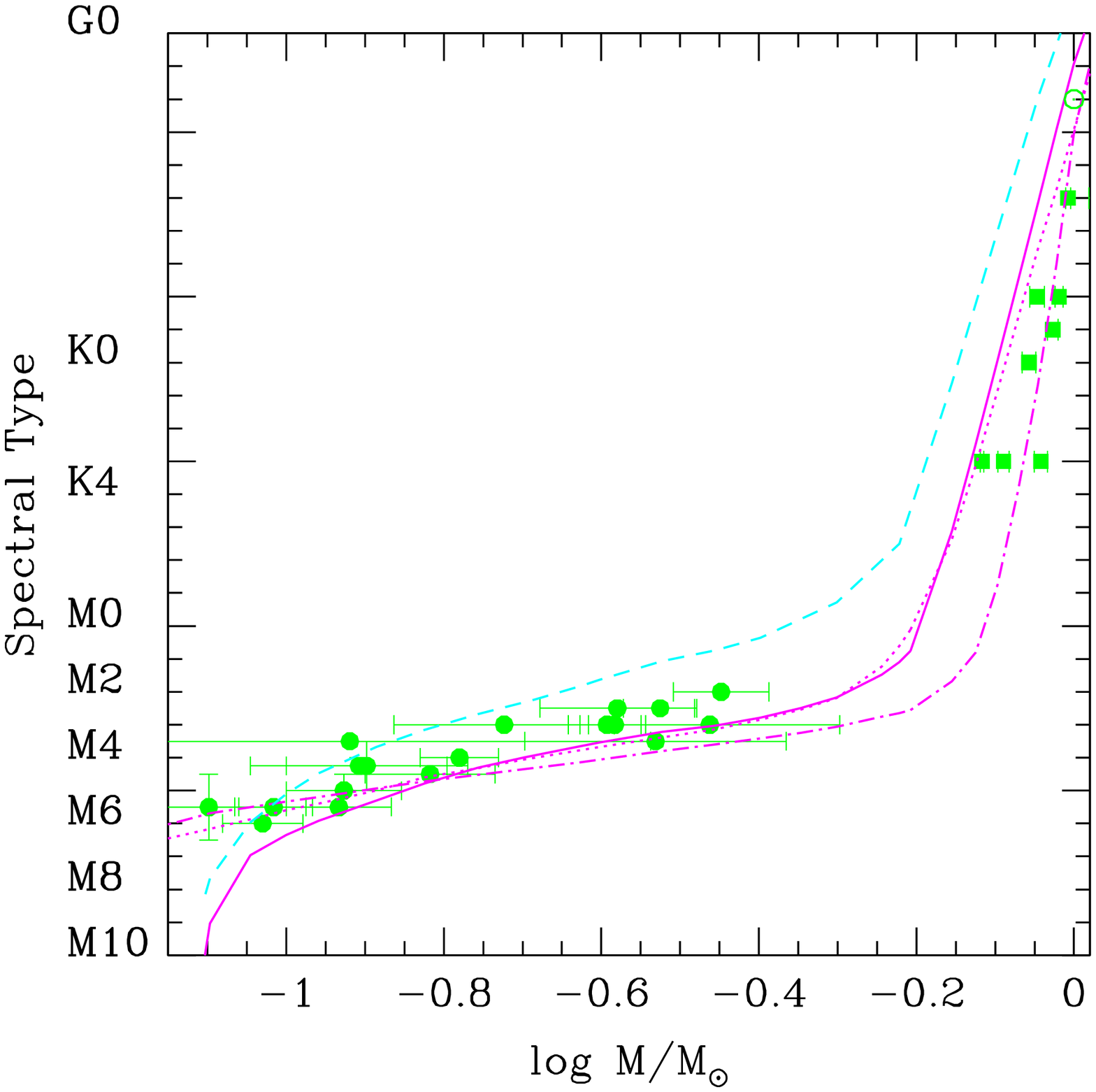}
\end{center}
\end{figure}

\vfill\eject

\begin{figure}
\begin{center}
\epsfxsize=180mm
\epsfysize=200mm
\epsfbox{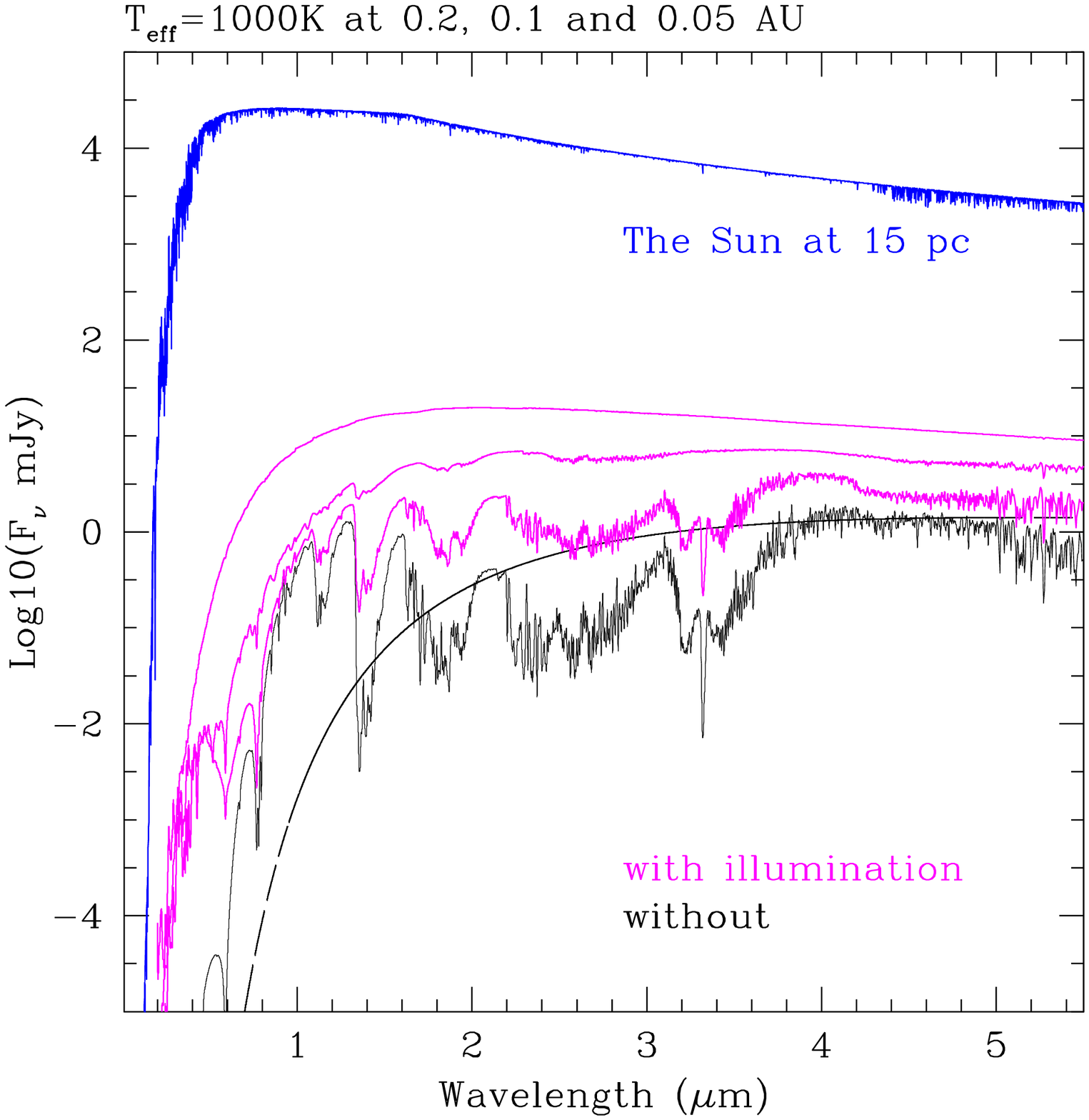}
\end{center}
\end{figure}

\vfill\eject

\begin{figure}
\begin{center}
\epsfxsize=180mm
\epsfysize=200mm
\epsfbox{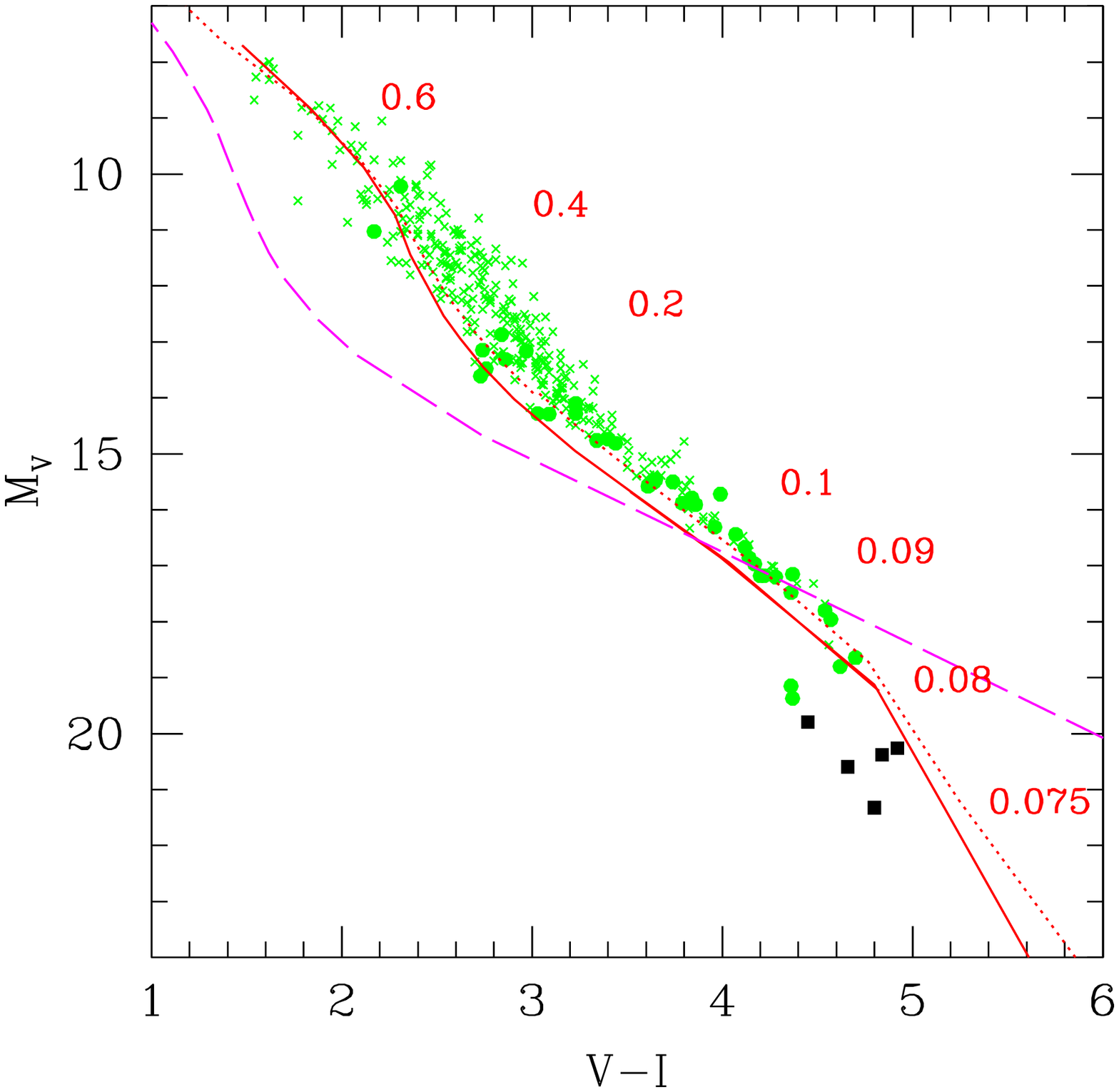}
\end{center}
\end{figure}

\vfill\eject

\begin{figure}
\begin{center}
\epsfxsize=180mm
\epsfysize=200mm
\epsfbox{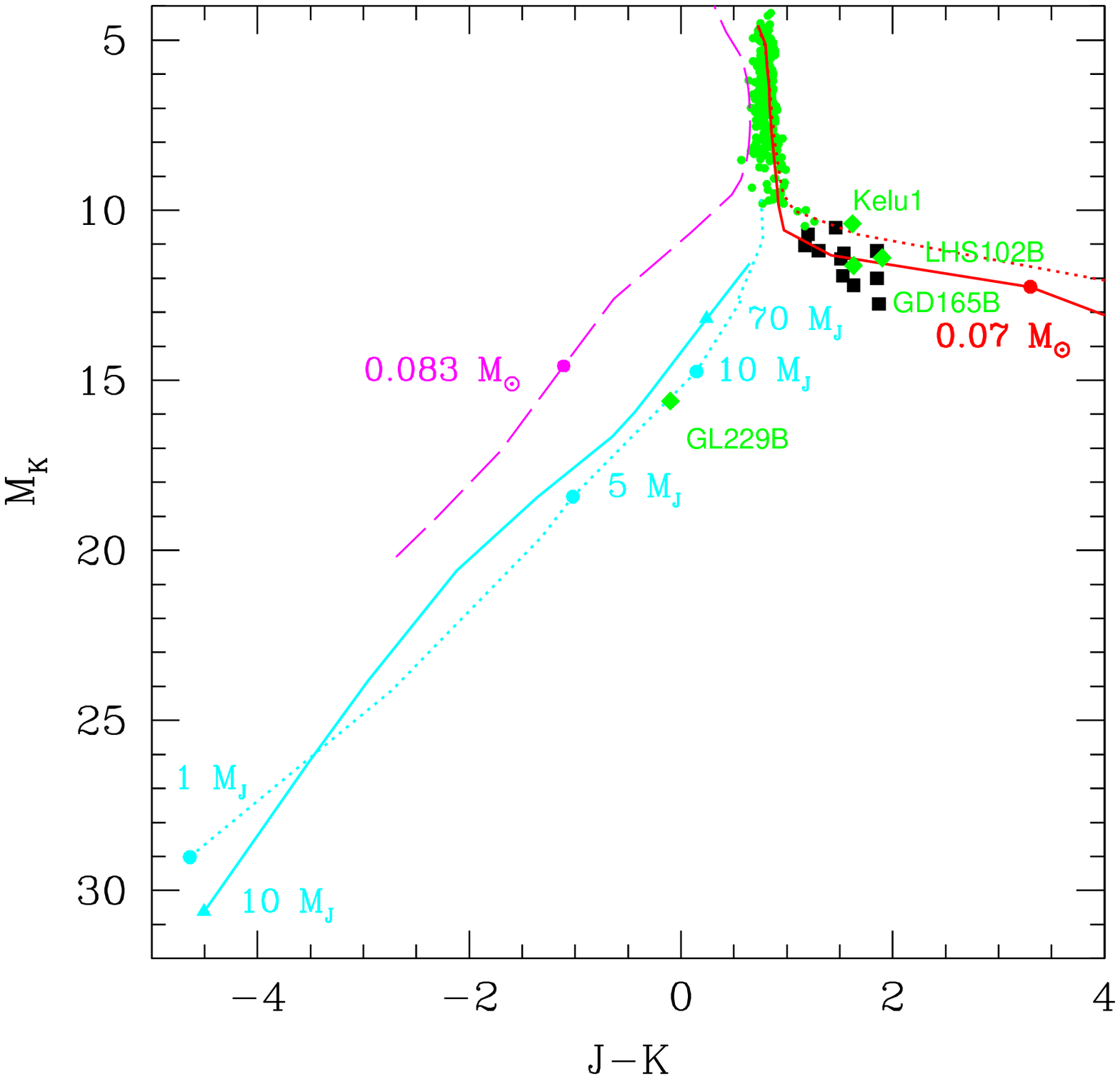}
\end{center}
\end{figure}

\vfill\eject

\begin{figure}
\begin{center}
\epsfxsize=180mm
\epsfysize=200mm
\epsfbox{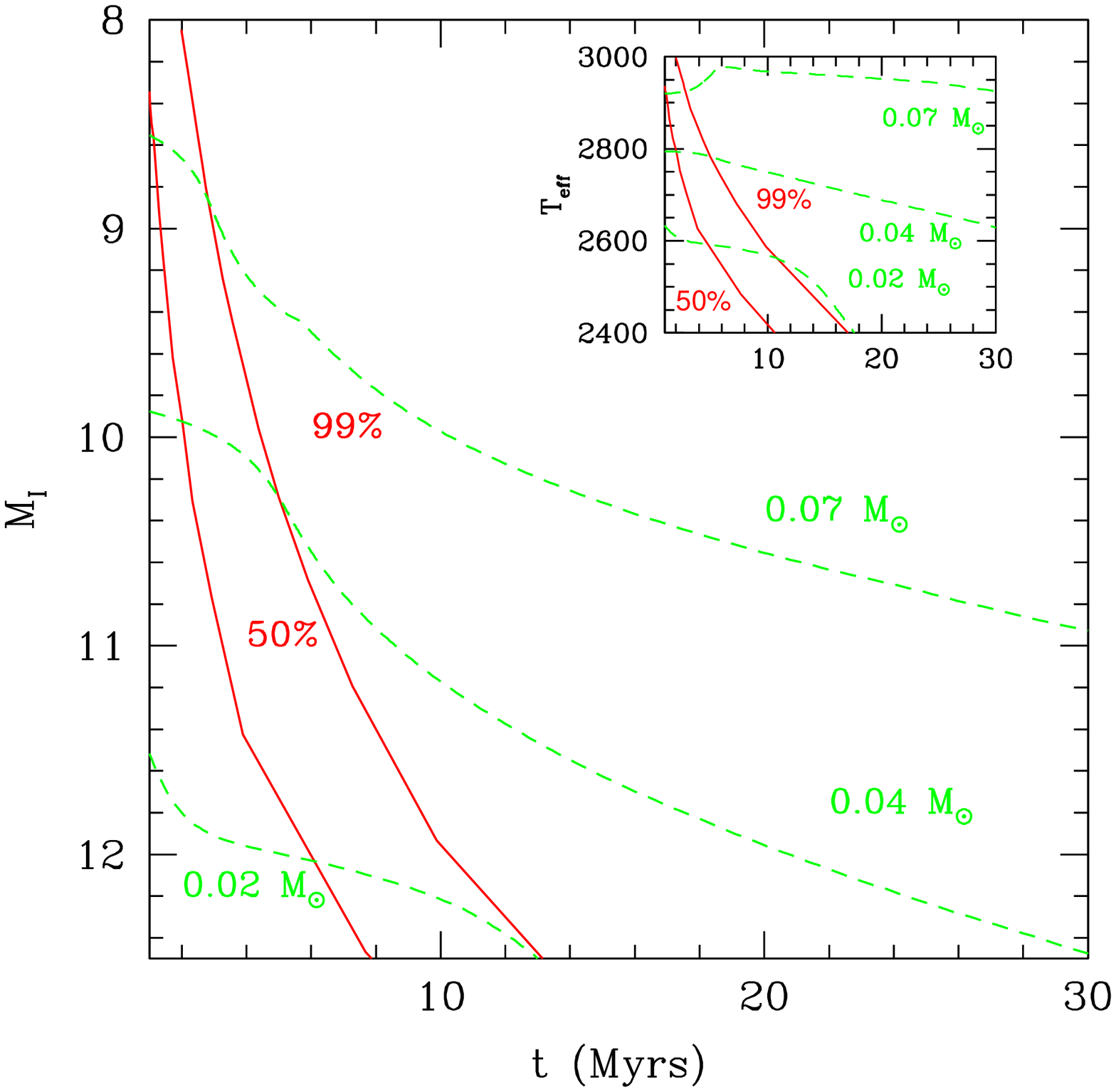}
\end{center}
\end{figure}

\vfill\eject

\begin{figure}
\begin{center}
\epsfxsize=180mm
\epsfysize=200mm
\epsfbox{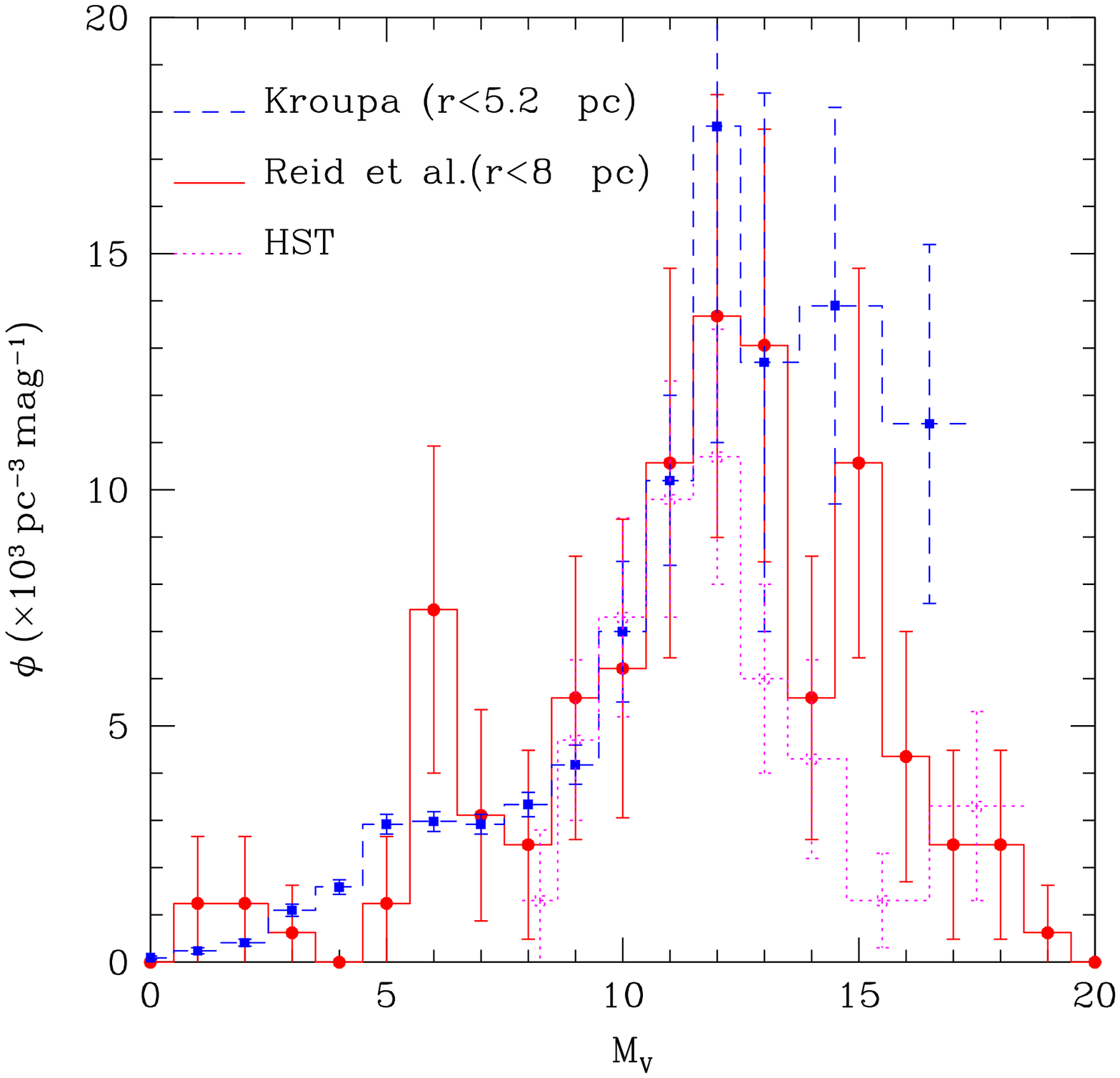}
\end{center}
\end{figure}

\vfill\eject

\begin{figure}
\begin{center}
\epsfxsize=180mm
\epsfysize=200mm
\epsfbox{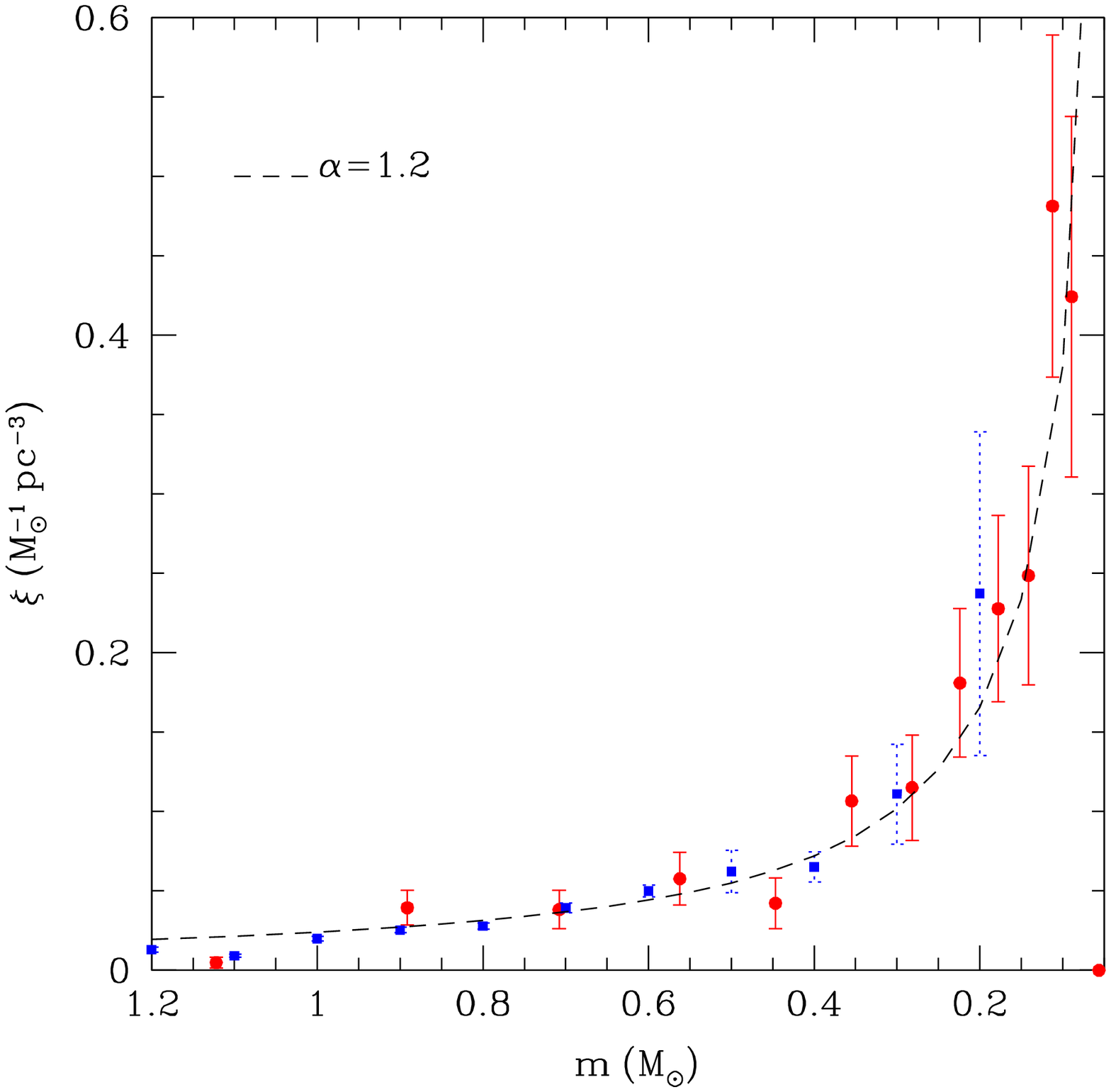}
\end{center}
\end{figure}

\vfill\eject

\end{document}